\def\LUM{\:{\rm ergs\:s^{-1}}}

\def\VEL{\:{\rm km\:s^{-1}}}

\def\OIGS{\:{\rm ergs\:cm^{-2}\:s^{-1}\:\AA^{-1}}}

\def\LB{Lyman\thinspace$\beta$}

\def\NiiiL{\ion{N}{3} $\lambda991$}
\def\NivL{\ion{N}{4} $\lambda923$}

\def\NvL{\ion{N}{5} $\lambda\lambda1239,1243$}

\def\CiiiLix{\ion{C}{3} $\lambda977$}
\def\CiiiLxi{\ion{C}{3} $\lambda1176$}
\def\CivL{\ion{C}{4} $\lambda\lambda1548,1551$}

\def\OviL{\ion{O}{6} $\lambda\lambda1032,1038$}

\def\SIivL{\ion{Si}{4} $\lambda\lambda1394,1403$}

\documentclass[12pt,preprint]{aastex}

\begin{document}


\newcommand{\MSOL}{\mbox{$\:M_{\sun}$}}  

\newcommand{\EXPN}[2]{\mbox{$#1\times 10^{#2}$}}
\newcommand{\EXPU}[3]{\mbox{\rm $#1 \times 10^{#2} \rm\:#3$}}  
\newcommand{\POW}[2]{\mbox{$\rm10^{#1}\rm\:#2$}}
\newcommand{\SING}[2]{#1$\thinspace \lambda $#2}
\newcommand{\MULT}[2]{#1$\thinspace \lambda \lambda $#2}
\newcommand{\CHINU}{\mbox{$\chi_{\nu}^2$}}
\newcommand{\vsini}{\mbox{$v\:\sin{(i)}$}}
\newcommand{\FUSE}{{\it FUSE}}
\newcommand{\HST}{{\it HST}}
\newcommand{\IUE}{{\it IUE}}
\newcommand{\EUVE}{{\it EUVE}}


\title{
The Effect of a Superoutburst on the White Dwarf  and Disk of VW Hydri as observed with \FUSE\
\footnote{Based on observations made with the NASA-CNES-CSA Far
Ultraviolet Spectroscopic Explorer. {\it FUSE} is operated for
NASA by the Johns Hopkins University under NASA contract
NAS5-32985}	   }

\author{Knox S. Long}
\affil{Space Telescope Science Institute, 3700 San Martin Drive, 
Baltimore, MD 21218}

\author{Boris T. G\"ansicke}
\affil{Department of Physics, University of Warwick,
Coventry CV4 9BU UK}

\author{Christian Knigge}
\affil{Department of Physics, University of Southampton,
Southampton SO17 1BJ UK}

\and

\author{Cynthia S. Froning}
\affil{Center for Astrophysics and Space Astronomy, University of Colorado, \\
593 UCB, Boulder, CO, 80309 USA}

\and

\author{Berto Monard}
\affil{Bronberg Observatory, CBA Pretoria, PO Box 11426, Tiegerpoort
0056, South Africa}

\begin{abstract}

We have used \FUSE\ to obtain a series of thirteen observations of the
nearby dwarf nova VW Hyi that cover the period from the end of a
superoutburst through the following normal outburst of the system.
Here, we present the quiescent spectra taken after each outburst
event. The spectra obtained during quiescence contain at least three
components. The dominant component over most of the \FUSE\ wavelength
range is the white dwarf (WD), which cools following the superoutburst.  The amount
of cooling is dependent on the white dwarf models utilized.  For log g
of 8.0, the temperature drops from 24,000K just after the outburst to
20,000 K just before the normal outburst.  For this model, and for a
distance of 65 pc, the radius of the white dwarf is approximately
\EXPU{8}{8}{cm}  and \vsini\
is $\sim 420 ~ \VEL$.  The fact that the derived radius is smaller than expected
for a WD with log g=8 suggests a higher gravity WD or that VW Hyi is somewhat 
further than its canonical distance of 65 pc.  Either is possible given the current uncertainty ($\pm$ 20 pc)  in the distance to VW Hyi.  Earlier suggestions that the WD photosphere
show evidence of CNO processed material are confirmed, but
our analysis also highlights the fact that significant issues remain
in terms of analyzing the spectra of WDs in such unusual physical
situations. The second component is relatively featureless and
shows substantial modulation on the orbital (and just after outburst, the superhump) period. The second component is most likely
associated with the hot spot where material from the secondary
encounters the disk, rather than emission from the boundary layer
region between the inner disk and WD.  This second component fades
about 10 days after the superoutburst.  There is also a third
component, clearly visible in terms of broad emission lines of
\CiiiLix, \NiiiL, and a combination of Lyman$\beta$ and \OviL , which
appears to be accompanied by a flat continuum.  The strength of the
emission lines, which are almost surely associated with the accretion
disk, appear relatively constant for the duration of the
observations. 

Subject Headings: stars: novae, cataclysmic variables -- stars:
individual (VW Hydri) -- stars: binaries: close -- ultraviolet: stars

\end{abstract}
\section{Introduction}
VW Hyi is a very short period SU UMa - type dwarf nova, which lies
along a line of sight with extremely low hydrogen column
density, possibly as low as N$_H$=\EXPU{6}{17}{cm^{-2}}
\citep{Polidan1990}.  As discussed by \cite{Smith2006}, and as is the
case with most cataclysmic variables, its system parameters are not
well-known.  The system inclination is generally taken to be 60$\pm$10
degrees, based on the existence of orbital humps and the lack of
eclipses \citep{Schoembs1981}.  The distance to VW Hyi is thought to
be about 65 pc, but this is based solely on the relationship
between period and outburst magnitude \citep[][see Sect. \ref{sec_wd}]{Warner1987}, despite the
fact that an astrometric determination should be straightforward
today. \cite{Smith2006} have carried out the most recent analysis of
the orbital parameters, and for the purpose of this analysis we will
follow their lead.  They estimate the mass of the white dwarf (WD) to be
0.71$^{+0.18}_{-0.26}$ \MSOL\ (based on the Sion et al.'s [1997]
determination of the $\gamma$ velocity of the WD and their
determination of the systemic velocity), which implies a WD radius of
\EXPU{9.3^{+2.5}_{-2.6}}{8}{cm} and a gravity of log g of
8.0$^{+0.38}_{-0.41}$, if the mass radius relationship for a C-O WD
applies.  This and the mass ratio of 0.148$\pm$0.08 \citep{Patterson1998a} imply a secondary mass
of 0.11$\pm$0.03 \MSOL, greater than that expected for a brown dwarf.
\cite[See][for an alternative view of the nature of the
secondary.]{Mennickent2004}

Like other members of the SU UMa group of dwarf novae, VW Hyi
periodically undergoes both normal outbursts and superoutbursts. 
Both types of outbursts occur as a result of a thermal
instability that transforms the disk surrounding the WD from a cold,
mostly neutral to a hot ionized state.  Outbursts occur fairly
regularly at intervals of about 20 days, raise the optical magnitude
from 14 to 9.5, and last about a day \cite[see, e.g.][]{Mohanty1995}.
Superoutbursts are typically separated by 150 days, last of order 12
days, and have a peak magnitude of 8.7.  

The actual trigger of superoutbursts is still debated, with the two
contenders being the thermal-tidal instability model \citep{Osaki1989}
and the enhanced mass transfer model \citep{Vogt1983}. Analyzing the UV
and EUV delay observed during superoutbursts of VW\,Hyi, as well as the
variability of precursor outbursts, \cite{Schreiber2004} concluded
that the enhanced mass transfer model provides a better agreement with
the observations.

VW Hyi was initially observed at FUV wavelengths with \IUE\ \cite[][and
references therein]{Gaensicke1996} and {\it Voyager}
\citep{Polidan1990} and has been observed subsequently with \HST\
\cite[][and references therein]{Sion2004}, the Hopkins Ultraviolet
Telescope \citep{Long1996} and, more recently, \FUSE\
\citep{Godon2004}.  In the high state, spectra of VW Hyi in the \IUE\  and
\HST\  range show
blue-shifted absorption lines of \NvL, \SIivL, and \CivL, and occasionally P-Cygni like profiles of C IV, superposed
on a fairly blue continuum.  The spectra, like those of other disk dominated cataclysmic variables, are understood as emission from the disk, modified by scattering in a wind emerging from the disk.  In quiescence, VW Hyi is dominated by
emission from the WD.  Outbursts appear to heat the WD surface, which
cools during quiescence. \cite{Gaensicke1996}, analyzing all of the
\IUE\ observations available, concluded that superoutbursts heat the WD
from about 19,300K to 26,400 K, and that the WD cools with a time
constant of about 9.8$^{+1.4}_{-1.1}$ days. By contrast,
\cite{Gaensicke1996} found that normal outbursts heat
the WD only to 23,800 K and that the cooling time constant in this
case is 2.8$^{+1.0}_{-0.5}$ days.  In quiescence, the flux near the Lyman
limit exceeds that expected from a uniform temperature WD
\citep{Long1996}.  \cite{Godon2004} conclude the second component in
the \FUSE\ spectrum resembles that expected from an ``accretion
belt'', but note its exact origin is uncertain.  \cite{Godon2005}
suggest that the emission arises from a region near the boundary layer
which also produces the X-ray emission that is seen from the system in
quiescence \citep{Pandel2003}.

We have obtained a series of thirteen observations of VW Hyi
undertaken with \FUSE\ beginning toward the end of a superoutburst
of VW Hyi and continuing through the next normal outburst of the
system.  The purpose of this series of observations, the
most intensive of any cataclysmic variable with \FUSE\  (or, for that
matter, any satellite since \IUE) was to better understand the effect
of outbursts in a cataclysmic variable on the WD and its implications
for the nature of the secondary component in the system.  Here we
describe an analysis of the spectra obtained in quiescence as well as associated
ground-based optical photometry; a detailed
analysis of the outburst spectra will be presented elsewhere.

\section{Observations and Reduction of the Data \label{Sec_Obs}}

Following notification by Janet Mattei of the AAVSO of the onset of a
superoutburst of VW Hyi, the \FUSE\ operations staff scheduled a
series of 13 observations of VW Hyi (labeled henceforth Observation
01--13). The onset of the outburst was not very well-determined; the
system was clearly in superoutburst on JD2453209.8 which triggered the
observations. By the time of the first observation, 7.4 days later,
the system had declined to m$_V$ of approximately 10.5.  A summary of
the \FUSE\ observations is given in Table
\ref{ObsLog}. The \FUSE\ campaign was accompanied by ground-based time
series photometry obtained at Bronberg Observatory (CBA Pretoria)
during 14 nights spanning the period JD2543221.42--3248.65 (labeled
henceforth Observation a--n). The optical data were obtained in white
light, using a 30\,cm Meade SCT\,LX200 with an SBIG ST7-XME CCD
camera. Data reduction and differential photometry was carried out
with AIP4WIN\,v.1. The spacing of the \FUSE\ and ground-based
observations, as well as the optical light curve for VW Hyi, as
compiled by the AAVSO, are shown in Fig.\,\ref{fig_aavso}.

A normal outburst of VW Hyi occurred on JD 2453232.7, about 16 days
after the system had returned to quiescence.  Obs.\ 11 occurred when
the system had a visual magnitude close to that of Obs.\ 1.  Obs.\ 13
occurred approximately 14 days after this normal outburst; the next
outburst appears to have occurred about two weeks later near JD2453259
(based a single observation of the system on that date and a number of
upper limits in the interim).

\FUSE\ contains a spectrograph with four independent optical channels
that together cover the wavelength range 905 to 1187 \AA.  The data
are recorded in eight segments with overlapping spectral ranges on two
separate photon-counting array detectors and are recorded as
individual photon events (TIMETAG mode) or as a two dimensional
images, summed over a period of order 600 seconds (HISTOGRAM mode).
TIMETAG mode is generally preferred, especially for faint sources,
because background subtraction in TIMETAG mode is more accurate and
because one can choose arbitrary time intervals for constructing
individual spectra.  All of the VW Hyi observations were obtained in
TIMETAG mode, with the exception of Obs.\ 2, which was inadvertently
obtained in HISTOGRAM mode (but during a time when VW Hyi was fairly
bright).  All of the VW Hyi observations were obtained through the
larger 30" x 30" LWRS aperture, which minimizes, but does not
eliminate, slit losses which occur primarily because of
thermally-induced misalignments of the optical channels.
\cite{Moos2000} and \cite{Sahnow2000} provide a description and
overview of the observatory and its performance.

For the analysis reported here, all of the data have been recalibrated
using version 3.2 of the ``calfuse'' pipeline \citep{Dixon2007}.  As shown
in Table \ref{ObsLog}, effective exposures times were relatively short
in the early observations (3000-6000 s in the the first seven
observations) and then were increased to 20,000-27,000 s in the last
six observations to reflect the anticipated decline in brightness of
the source as a function of time from outburst.  There were no major
anomalies in any of the observations, except the high voltage of the
Side 1 detector was turned off for much of Obs.\ 7 (probably as a
result of a single event upset).  The standard pipeline products are 8
spectra, one for each optical channel/detector combination, created in intervals chosen by
the pipeline.  For the analysis which follows, we have constructed
average spectra for each observation from these spectra.  To account
for slit losses, we have renormalized fluxes from the individual
spectra so that the average flux near 1060 \AA\ is the same as that of
the LiF1B channel, which was used for guiding.  In constructing the
average spectra, we weighted each data point by the product of the
effective area and effective exposure time for that data point. We
also used visual inspection to trim the wavelength ranges used for
summing the individual channels. Except where noted otherwise the
spectra are rebinned from the native wavelength grid (with a
resolution of about 0.01 \AA), to a resolution of 0.1 \AA.

As indicated in Fig.\ \ref{fig_spectra}, the spectra show an evolution
with time.  In particular the fluxes decline at all wavelengths from
Obs.\ 1 through Obs.\ 10, the last spectrum before the normal outburst.
The outburst spectrum obtained during Obs.\ 11 (not shown in Fig.\
\ref{fig_spectra}) is almost identical both in shape and in flux to
that in Obs.\ 1.  The Obs.\ 12 spectrum obtained about 3 days after
the normal outburst has a flux level similar to the last spectrum
prior to the outburst, and the last spectrum obtained in Obs.\ 13 has
the lowest flux of all.  Declines in flux are accompanied by an
overall ``softening'' of the spectrum, with the fluxes declining more
rapidly at 950 \AA\ than at 1150 \AA.  There appears to be continuum
emission down to the Lyman limit in all of the observations, but it is
less evident far from outburst than just after outburst.

Not surprisingly, the character of the quiescent spectra are
significantly different than those obtained at the end of
superoutburst (and during the normal outburst).  The Obs.\ 1 (and 11)
spectra show absorption features due most obviously to \NivL,
\CiiiLix, and \NiiiL, which are not seen in the short wavelength
portion of the quiescent spectra.  The \LB\ profile is broader in the
quiescent spectra, and the quiescent spectra tend to show \OviL\ in
emission, versus \OviL\ in absorption in outburst.  The emission
profiles seen in quiescence are centered near zero velocity and are
most likely due to emission from the disk, as is the case WX Hyi and SS
Cyg \citep{Long2005}. The quiescent spectra also show far narrower and
deeper absorption features than seen in outburst.  All this is
consistent with the hypothesis that the quiescent spectra are due
primarily to emission from a fairly slowly rotating WD, while the Obs.\
1 (and 11) spectra are dominated by an accretion disk.

\section{Short-term UV and optical variability \label{Sec_Var}}

Our initial inspection of the \FUSE\ spectra obtained from the
standard data processing did not reveal major variations in the
emission properties of VW Hyi within individual observations.
However, to investigate the ultraviolet time-variability properties
more quantitatively, we created time-resolved spectra and light curves
using software tools provided by the FUSE Project at Johns Hopkins
University, and described in the ``IDF
Cookbook'' \citep{Godard2004}. These data represent the first
opportunity to investigate in some detail the short-term variability
of VW Hyi at ultraviolet wavelengths.  

In the following discussion, we assume that VW Hyi has an orbital period 
$P_\mathrm{orb}$ of $106.950295(20)$\,min \citep{VanAmerongen1987}. The
accumulated error in the ephemeris of \citet{VanAmerongen1987} is
$\simeq0.033$, which is negligibly small in the context of this
paper. Following \citet{Smith2006}, we apply an offset of 0.15 to the
zero-point of the \citealt{VanAmerongen1987}, so that phase zero
corresponds to the inferior conjunction of the secondary star, 
\begin{equation}
HJD_\mathrm{max}=2\,440\,128.03521(59)\pm0.074271038(14),
\label{e-eph}
\end{equation}
i.e. maximum light of the orbital modulation occurs at phase 0.85.

\subsection{Optical data}

In a first explorative step, we subjected the initial optical data
sets, as well as various combinations of mean-subtracted data sets to
an \texttt{ANOVA} time-series analysis, which is particularly
sensitive to the detection of periodic, but non-sinusoidal
variability \citep{Schwarzenberg1996}. The morphology of the
variability clearly changed between the observations taken before the
next normal outburst, and those taken after it. We therefore decided
to analyse the two periods separately.

The ANOVA periodogram computed from the combined observations b--h
(Figure\,\ref{tsa_berto}, top left panel) shows a 1-day alias pattern
with maximum power near $\simeq109.5$\,min, i.e. somewhat longer than
the orbital period of VW\,Hyi of \citet{VanAmerongen1987}. Fitting a
multi-harmonic sine wave to the optical data results in
$P_\mathrm{sh}=109.4753(76)$\,min, and we identify this modulation as
the superhump period. Phase-folding the optical data with that period
results in a light curve with relatively large scatter
(Figure\,\ref{tsa_berto}, top right panel).

The periodogram calculated from the data pre-whitened with the period
obtained from the sine fit above contains again a 1-day alias pattern,
now peaking near $\sim107$\,min (Fig.\,\ref{tsa_berto}, panel (b)). A
multi-harmonic sine-fit to the detrended data results in
$P_\mathrm{orb}=106.9312(51)$\,min. This value is very close to the
the orbital period of VW\,Hyi measured by \citet{VanAmerongen1987},
though not formally consistent. \citet{VanAmerongen1987} determined the orbital period by fitting a linear ephemeris to times of maximum light in the orbital variation of VWHyi spanning one decade, and did not note any evidence for a period change. The formal discrepancy (3.7$\sigma$) between their orbital period and ours is most likely due to an underestimate in the error of our orbital period,  based on sine-fitting relatively sparse data. The light curve obtained from folding the
optical data pre-whitened with $P_\mathrm{sh}$ on the ephemeris
of \citet{VanAmerongen1987} exhibits a double-humped morphology
(Fig.\,\ref{tsa_berto}, right-hand-side panel (b)), as observed in
many short-period dwarf novae (e.g. \citealt{Patterson1998,
Rogoziecki2001, Araujo2005}). Maximum brightness occurs near phase 
0.85, as expected according to Eq.(\ref{e-eph}) for an orbital
modulation.

For completeness, we also computed a ANOVA periodogram for the
combined data set detrended with the orbital period, which shows, as
expected, a clear 1-day aliases pattern centered on the superhump
period (Fig.\,\ref{tsa_berto}, panel (c)). Folding the optical data,
pre-whitened by $P_\mathrm{orb}$, on the superhump period displays a
smooth quasi-sinusoidal modulation. 

The first optical data set is worth an additional note. Its power
spectrum is clearly dominated by the superhump modulation. However,
the superhump modulation underwent a phase shift and/or slight period
change between observations a and b, which has been observed
previously in VW\,Hyi and other dwarf novae
(e.g. \citealt{VanDerWoerd1987}), and we hence did not include
observation a in the combined data set.

Observations j--n were obtained following the next normal outburst,
and we combined them into a single data set for subsequent time-series
analysis following the same procedure as above. The ANOVA periodogram
computed from these data (Fig.\,\ref{tsa_berto}, panel (d)) contains
an alias pattern with the highest power at $\simeq107$\,min. From a
multi-harmonic sine-fit to the data we obtain
$P_\mathrm{orb}=106.9421(35)$, i.e. identifying an orbital
modulation. The orbital phase-folded light curve, using
Eq.\,(\ref{e-eph}), has a similar double-humped morphology to that
detected in observations b--h. Maximum brightness occurs again, as
expected, around phase 0.85. A periodogram computed from the data
pre-whitened with the multi-harmonic sine-fit at the orbital period
exhibit no further significant signal.

We conclude that the variability seen in the optical observations
following the superoutburst, but preceding the next normal outburst,
is dominated by a superhump modulation, but a clear orbital variability
is present as well. Following the next normal outburst, the optical
light curve of VW\,Hyi only exhibits an orbital modulation. 

\subsection{\FUSE\ data}

As outlined for the optical data above, we first analyzed the
individual mean-subtracted \FUSE\ light curves, as well as various combinations of
individual data sets. Similar to the behavior seen in the optical
data, a change in the morphology is observed between the \FUSE\ data
obtained early after the end of the superoutburst, and data taken
somewhat later. 

To analyse the variability shortly after the end of the superoutburst,
we combined the \FUSE\ observations 03--05 (Table\,\ref{ObsLog},
Fig.\,\ref{fig_aavso}). An ANOVA periodogram calculated from these
data (Fig.\,\ref{tsa_fuse}, panel (a)) shows maximum power near
$\simeq109.9$\,min. A multi-harmonic sine fit to the data results in
$P_\mathrm{sh}=109.866(43)$\,min, which is marginally longer than the
superhump period found from the optical observations. However, as
mentioned above, the superhump period and/or phase may drift slightly
shortly after the end of the superoutburst.

Next, we combined the \FUSE\ observations 08--10. The periodogram
computed from these data (Fig.\,\ref{tsa_fuse}, panel (b)) shows an
alias pattern with maximum power at $\simeq106.9$\,min, refined to
$P_\mathrm{orb}=106.8836(82)$\,min from a multi-harmonic sine
fit. Folding the data on Eq.\,(\ref{e-eph}) produces a double-humped
light curve with maximum brightness near phase 0.85, very similar to
the orbital modulation seen at optical wavelengths.

A time-series of the \FUSE\ observations 12 and 13, taken after the
next normal outburst, reveals some marginal power near the orbital
period, but the signal-to-noise ratio is too low for any detailed
analysis. 

We conclude that the variability at ultraviolet wavelengths follows a
similar trend as that observed in the optical: whereas the
early \FUSE\ observations seem to be modulated at the superhump
period, the observations taken a few days later are clearly modulated
at the orbital period. This implies that we are witnessing an flux
component in addition to the emission from the white dwarf which
exhibits a strong orbital phase dependence. This is illustrated in
Fig. \ref{fig_lightcurve}, which shows the wavelength averaged flux
from VW Hyi as a function of the number of orbital cycles elapsed
since JD 2453217.11732.  This time, which is just prior to the
beginning of our first observation, was selected so that integral
periods correspond to phase 0, given the ephemeris in
Eq.\,(\ref{e-eph}) corresponding to the inferior conjunction of
secondary conjunction, and approximately 0.15 cycles after the maximum
brightness of the orbital modulation.  There is excess emission in the
phase range 0.6 to 1, corresponding to hot spot maximum in these
two observations.  Similar variability is seen in earlier
observations, although the earlier observations cover fewer orbital
periods, which makes it harder to identify orbital trends with
confidence. The variations are less prominent in Obs.\ 10, and not
apparent in Obs. 12 and 13 in the lightcurves themselves.  Except in
Obs.\ 5, peak fluxes are observed near phase 0, that is at hot-spot
maximum.  The outburst spectra do not show this kind of variability.

Spectra from the high and low flux phases of the Obs.\ 8 and 9 are
shown in Fig.\,\ref{fig_phase_resolved_spectra}.  The spectra were
created by using the tool ``idf\_cut'' \citep{Godard2004} which allows
one to construct the average spectrum in a phase interval for an
entire observation.  In this case, we split the data into intervals
corresponding to each eighth of a phase. This yielded a total of eight
spectra, each corresponding to time-averaged spectra for each eighth
of a phase.  We then constructed the ``high flux'' and ``low flux'
spectra by averaging the three highest flux and the lowest flux
spectra, so that ``high flux'' and ``low flux'' spectra each contain
all of the data obtained over three-eighths of a phase.  Specifically,
the ``high flux'' and ``low flux'' spectra include data from phases
0.625-0.0 and 0.0-0.375, respectively.  This method is preferred
to splitting the data into a larger number of segments, constructing
spectra of the individual segments, and then combining them because,
with longer individual segments, the \FUSE\ software is able to model
the background.

Inspection of Fig.\,\ref{fig_phase_resolved_spectra} makes it clear
that there is a variable continuum component that is much stronger in
the ``high flux'' than in the ``low flux'' spectrum; the difference
spectra are shown in the lower panel of
Fig.\,\ref{fig_phase_resolved_spectra}. The general shapes of the
difference spectra are similar in Obs.\ 8 and 9, and in the other
observations where phase-dependent flux modulations are observed. This
second component accounts for much of the flux below 950 \AA, and
fills in the flux near \LB\ in the ``high flux'' spectra. That said,
there is no phase interval in which the flux at the center of \LB\
falls completely to zero, as would be expected for a pure WD. This
suggests that the ``second" component always contributes to some
extent (or, more precisely, that there is {\em some} non-WD component
present all of the time).  In this regard it is interesting that the
spectrum obtained during the low flux phases shows broad O VI in
emission; this is usually interpreted as emission from the disk. The
second component is not completely smooth, with some evidence of broad
absorption at \LB\ and a few higher order Lyman lines, in addition to
C III, which is clearly in absorption.  There are no emission features
apparent in the second component spectrum.

\section{Spectral Analysis}

In order to characterize the spectra obtained of VW Hyi in the
quiescent interval, we carried out a series of model fits to the data
using a large grids of simulated WD spectra calculated with Ivan
Hubeny's TLUSTY and SYNSPEC suite of programs
\citep{Hubeny1988,Hubeny1995}.  Our initial grid, which we have used
previously for modelling WZ Sge and U Gem, consisted of model spectra
ranging in temperature (T$_{wd}$) from 10,000 K to 150,000 K (in units
of 1000 K below 30,000 K).  The grid contains models for atmospheres
whose photospheric metal (and He) abundances range from z of 0.01
solar to 10 solar.  The grid contains model spectra appropriate for WD
rotation rates (\vsini) ranging from no rotation to rotation rates of
1200 $\VEL$ (in steps of 25 $\VEL$ for the range of relevance for VW
Hyi).  The grid contains models with gravities ranging from log g=7.5
to 9.0, in steps of 0.5.  However, we do not typically attempt to fit
the gravity, because there is a strong temperature-gravity degeneracy
in the models: one can almost always find a somewhat higher
temperature, higher gravity model that fits the data about as well as
a given lower temperature, lower gravity model (and vice versa). This
is due to higher gravities resulting in stronger Stark broadening of
the hydrogen lines, which is compensated by raising the ionization
fractions through an increase in temperature. 
For VW Hyi, the gravity is likely to be around 8.0 \citep{Smith2006},
and, unless otherwise noted, the models we will discuss are for
this gravity. Where relevant, we will indicate how the results would
change for other gravities.

We used a standard \CHINU\ fitting routine and fit the entire
spectrum, except for those regions near strong interstellar absorption
lines and air glow lines.  We used the same set of fixed set of
wavelength intervals for all of the fits.  As is common in our
experience with high-S/N spectra obtained with \FUSE\ (even with
well-understood systems), our fits yielded \CHINU\ that are typically
larger than 1. This makes it impossible to directly estimate
confidence intervals on the fit parameters via the usual
$\Delta\chi^2$\ method.  There is no straightforward way to deal with
this, as the causes of the problem range from poorly understood
systematics in instrumental sensitivity or background subtraction, to,
in our case, inaccuracies in the atomic parameters that
determine the strength of individual lines.

In order to get a handle on the likely uncertainties associated with
our fits, we have elected to characterize the errors using the 
bootstrapping method, see e. g. \cite{NumRecipes}.  Briefly, in the
bootstrapping method, one fits multiple versions of the data.  If the
original version of the data set contains N data points, one randomly
selects N data points to fit, allowing individual points to be sampled
multiple times to form the new version of the data. One then uses the
dispersion in the results to estimate the errors for the parameters in
the model.  Unless otherwise noted, the errors we quote are 1 $\sigma$,
that is 67\% of the trials gave parameter values within the quoted
ranges. Real features in the spectra and the models typically have
a width of about 3 \AA, and hence while the data points within such
an interval are formally independent of one another in a statistical
sense, any systematic errors in the models (for example due to
incorrect atomic data), will affect a region of at at least this width.
Therefore, in our creation of the bootstrap sample, when we pick a
data point for inclusion, we also include all of the "good" data
points within 1.5 \AA\ of the original data point.

\subsection{The White Dwarf}
\label{sec_wd}

It should be clear from the discussion of the phase-dependent time
variability in Sec.\ \ref{Sec_Obs} that spectral modeling of the VW
Hyi observations will be complex.  
In general terms, the process we followed was as follows. In fitting
the data, we assumed that there is no reddening along the line of sight
to VW Hyi, but we did allow for interstellar absorption lines from H
and O I, assuming a H column density corresponding to log N$_H$ =
18.6\footnote{This value is higher than deduced by \cite{Polidan1990}
from \IUE\ data and reflects our preliminary curve-of-growth
analysis of the column density. Since essentially all of these lines
lie in regions excluded from the model fits, \CHINU\ is not affected
significantly by this choice. We will present our N$_H$\ determination
more fully in a future paper discussing the high-state spectra.}

Primarily to establish a baseline, we began with simple uniform
temperature WD models.  These models have three variables, T$_{wd}$,
the overall metal abundance z, \vsini, and the normalization of the
spectrum.  The normalization is proportional to the effective solid
angle.  These models do qualitatively fit many aspects of the data.
As expected for a WD with T$_{wd} \simeq 20,000$\ K, the models match
the general shape of the spectrum, show broad absorption from \LB\,
and replicate, at least in a general way, many of the line features in
the spectrum.  For log g=8 models, the derived T$_{wd}$ drops from
27,000 K to 20,000 K from Obs.\ 2 (\CHINU=10.0) to Obs.\ 13
(\CHINU=5.4).  For Obs.\ 8, the ``low flux'' spectrum, that is data
between phase 0.0 and 0.375 yields a T$_{wd}$ of 22,000 K
with \CHINU=3.7, compared to a value of 24,600K with \CHINU=5.6 for
the ``high flux'' spectrum from phases 0.625-1.0.

There are several problems with this ``baseline model'', but the obvious one
is that it does not account at all for the flux from any extra components, of which the time-variable one is the most obvious.
One way to attack this problem is to model the second component as
a simple power law, a blackbody, or, as is commonly done, a second synthetic stellar spectrum.  In reality, as we have already indicated that there must be at least 3 components, the WD, one associated with the hotspot, and a third associated with the broad emission lines, and so any-two component approach will have its own limitations.  In principle we could have added the emission lines explicitly. In practice we have excluded the regions where the emission lines are obvious from the fits.

We have tried all three two-component approaches.   As an example of this approach, we describe
model fits of  the data involving 
combinations of emission from the WD surface and a second component
modeled as a simple power law, e. g.  \begin{equation} \rm
f_{\lambda}(\lambda) = f_{\lambda} (1000 \AA) ( \lambda/1000
\AA)^{\alpha-2}.  \end{equation} (With this formulation, $f_\nu
\propto \nu^{-\alpha}$.)

This produced much better fits than the simpler one-component WD models, both qualitatively and in terms of
\CHINU\ for most of the spectra.  The results of log g=8.0,
scaled-solar abundance, WD plus power law models are summarized in
Table \ref{wd80pow} for the time-averaged spectra and for the
phase-resolved spectra of Obs.\ 8 and 9. The best fit for Obs.\ 2 now
has T$_{wd}$ of 24,300K and \CHINU\ of 3.6 (compared to 10
previously), while for Obs.\ 13, T$_{wd}$ is 19,500K and \CHINU\ is
4.2 (compared to 5.4).  The combination of WD and power law components
produces a much better approximation to the observed spectra, both in
terms of fitting the shape of the spectrum near \LB\ as well as
providing the correct fluxes near the Lyman limit.  Compared to the
baseline, pure WD model, the implied values of T$_{wd}$ are smaller by
1,000-2,000 K, and the scaled-solar abundances are larger. Both of
these differences are easily understood. The power law component
dominates the fitted spectrum near the Lyman limit, and therefore the
WD tends to be cooler.  The power law component also dilutes the WD
portion of the spectrum somewhat even at the longer wavelengths, where
the metal lines are most prominent. This is compensated in the fits by
increasing the metallicity of the WD component. However, despite these
relatively minor differences, the same general trends are still
observed as a function of time from outburst, i.e. a cooling of the WD
and a gradual decrease in metallicity.  Note the the fit including the
second component is not only better for the high-flux spectrum from
Obs.\ 8 (\CHINU\ is 2.6, compared to 5.6), but also for the low-flux
spectrum (\CHINU\ is 3.1, compared to 3.7), confirming that there is
evidence for a second (or third) component even in the low-flux
spectrum.

An example of a WD plus power-law fit, the fit to the time-averaged
spectrum of Obs.\ 8, is shown in the upper panel of Fig.\
\ref{fig_vwhyi8_metal_var}.  The quality of this fit is fairly
typical.  The overall shape of the spectrum is fit well, and most, if
not all, of the features in the data are seen in the model, although
not always at the correct strength.  The effect of the continuum
component is plainly seen in the core of \LB\ and near the Lyman
limit.  There are some troublesome regions, however, especially near 1140
\AA, where there is a strong feature in the models, which is not
present in the data, and also near 1128 \AA, where there is a strong
feature in the data, which is not present in the model.  The 1140 \AA\
feature in the models is due to C II, as one can easily determine by
constructing models with lower C abundances while holding other
elements fixed.  It seems natural to associate the feature at 1128
\AA\ with P V, except that there seems to be little evidence of the
stronger component of the doublet at 1118 \AA. Over and above these 
specific difficulties, it is clear that some of the line shapes are
not especially well fit. One way to see this is to examine the
``errors'' on \vsini, which are quite substantial in our bootstrapping
approach. This basically reflects the fact that the line widths are
not consistent with a single \vsini\ using this model.

The WD plus power-law, WD plus BB, and WD plus stellar atmosphere parameterizations of the
quiescent spectra of VW Hyi all result in qualitative and quantitative
improvements to the fits, and they do so mainly by fitting the overall
spectral shape.  In terms of the primary WD component, the results are almost identical, and from a statistical (\CHINU) perspective, there is no reason to choose one or the parameterizations over another. We have also carried out similar fits using WD models
with different gravities.  Again, the results are very similar. The
main difference is that best fit temperatures are somewhat higher for
higher gravity models.  This is a direct result of the fact that at
fixed temperature, the Lyman lines deepen as the gravity is increased,
but at fixed gravity, they weaken as the temperature is increased.
 
In order to address some of the problems that remained, we next
considered models in which the abundances of various elements in the WD were
varied independently.  The strongest absorption lines in the spectra
are lines of C, N, Si and S, so we created grids of WD models with varying
abundances for these elements.  We then again fit the observed
spectra with a WD plus power-law (or WD plus BB) model. The results were
very similar in both cases, and are shown for the log g=8 WD plus 
power-law model in Table \ref{wd80abun}.  The improvement in the
fit can be seen by comparing the fit to the time-averaged spectrum
from Obs.\ 8 in the lower panel of Fig.\ \ref{fig_vwhyi8_metal_var} to the
fit shown in the upper panel, in which only the overall metallicity is
varied.  The models with variable abundances fit the data better in the
long wavelength portion of the spectrum (which is where the strongest
absorption lines are located and the S/N is highest).  Specifically, they
improve the fits to Si III complexes near 1110 \AA\ and 1140 \AA.  One
measure of the improved fits to the lines is that the errors on \vsini\
have dropped by nearly an order of magnitude to 30 $\VEL$, or about
one-tenth of the value of \vsini\ itself.

The successes and deficiencies of the models appear to be fairly
universal, as a comparison of the fits to the time-averaged spectra
from Obs.\ 2 and 13, shown in Fig.\ \ref{fig_vwhyi_abun}, reveal. The
fitted value of T$_{wd} $ is 24,100 K for Obs.\ 2, and 19,000 K for
Obs.\ 13.  Both fits require a second component, although the flux at
1000 \AA\ implied by the power law component is an order of magnitude
higher immediately after the outburst than in Obs.\ 13, where there
was no evidence of phase-dependent variations.  (The underlying shape
of the continuum looks a little different, with more emission
components near the Lyman limit, near CIII 978, NIII 990, \LB, and O
VI in the Obs.\ 13 spectrum.) Both fits describe the overall spectrum
fairly well, and the main problem areas have also remained fairly
consistent (e.g. near 1195 \AA, near 1128 \AA, and near 1140 \AA).

The improvements in the fits are similar regardless of whether one is
considering the ``high flux'' or ``low flux'' spectrum of Obs. 8, as shown in
Fig. \ref{fig_vwhyi8_phase_abun}. This comparison also reveals a
concern, however: in both Obs.\ 8 and 9, the inferred T$_{wd}$ is hotter
in the spectrum from the high-flux state than that from the low flux
version.  This may indicate that our description for the second
component is too simplistic. If so, the difference between the high-
and low-state WD temperatures ($\simeq$\ 1000~K) is an estimate of the
systematic error associated with our ignorance regarding the nature of
the second component. 

The model fits indicate that the abundance of C is about half solar,
regardless of time from outburst.  A smaller C abundance not only fits
the C III lines better, but also reduces the intensity of of a C II
feature near 1140 \AA\ that has a substantial EW at higher abundances.
The abundances of N and Si all appear to be greater than solar in
these model fits, whereas S at least after the first week after the
end of superoutburst is near solar.  There still appears to be secular trend toward
lower metallicity as a function of time from outburst, but it is
weaker than for the scaled-solar abundance plus powerlaw fits.  

The abundance patterns that are observed in the \FUSE\ spectra of VW Hyi are typical of those expected in
CNO processed material.  Similar N and C abundances have been reported
in VW Hyi using \HST\ \citep{Sion2001}.  Both sets of spectra
include \CiiiLxi, which is important in constraining the C abundance,
but the N features in the \FUSE\ spectra, most
notably \ion{N}{2}$\lambda$1085, are disjoint from those observable
with \HST.\footnote{\cite{Sion2001} also reported a large overabundance of P  in their analysis 
of  \HST\ spectra, which were obtained in early quiescence following a superoutburst, which is interesting in view of the discussion concerning the P V doublet.  While our observations were obtained at a similar time in its outburst cycle, we still feel that it is unlikely that the 1128 \AA\ feature is due to P V, both because of the absence of other half of the doublet, and because we have seen the same problem in modeling \FUSE\ spectra of other systems, most notably U Gem \citep{Long2006}.  }
Similar abundance patterns have been reported from spectroscopic analyses of the WDs in some other CVs, including U Gem \citep{Long2006}, and with less precision from analyses of emission line spectra of both magnetic and non-magnetic systems \citep{Gaensicke2003}.  The carbon abundances of the secondary stars in a number of dwarf novae (but not magnetic CVs) appears to be very low, based on the weakness of the CO bands in the IR spectra \citep{Harrison2005a,Harrison2005b}.   The photospheric abundances on the WDs of DNe should reflect that of accreted material from the disk, since metals in a WD quickly sink through the photosphere \citep{Paquette1986}.  Indeed, photospheric abundances in the WDs of CVs far from outburst tend to be less than solar, which is thought to reflect the rapid diffusion of accreted material into the interior \citep{Gaensicke2005}.  The C abundance of the secondary in VW Hyi has not yet been measured, but one would expect to find that it is also low.  The XMM grating spectra of VW Hyi show emission from N VII, but do not extend to long enough wavelengths to include the C lines. \cite{Pandel2003} carried out variable abundance fits to the spectra in the context of a cooling flow; all of the abundances they derived were close to or consistent with solar.  Several explanations have been proposed to explain CNO enhanced abundances observed in dwarf novae, including the nuclear evolution of the a secondary star,  transfer of CNO material from the primary during the common envelope phase, or pollution of the  secondary star with material from nova explosions \citep{Marks1998}.  \citep[See the discussion for VW Hyi in particular by][]{Sion2001} The actual cause has not been resolved although the absence of CNO enhancements in the secondary stars of pre-cataclysmic binaries appear to favor the nova hypothesis \citep{Tappert2007}.

Despite the residual problems in the model fits, we believe that the
variable abundance models provide a more accurate 
representation of the system as a function of time than do the simpler
models.  The apparent trends in T$_{wd}$, R$_{wd}$, \vsini, and the
contribution of the second component are shown in Fig.\
\ref{fig_trends} for log g=7.5, 8 and 8.5 models in which the second
component is a power law.  The trends are similar when the second
component is modelled as a blackbody or a stellar atmosphere. As was the case for the
scaled-solar abundance fits, 
the temperatures for log g=8 are lower by about 2000K than for log
g=8.5, which causes the apparent radius for log g=8 to be about 25\%
higher than when log g=8.5 is assumed.  For fixed gravity, 
the derived radius is fairly constant with time. 

Based on the standard \cite{hamada1961} relationship, the expected
radius for a ``cold'' WD is \EXPN{11.9}{8}, \EXPN{8.9}{8}
and \EXPU{6.3}{8}{cm} for log g=7.5, 8, and 8.5 respectively, while the more
realistic \cite{Bergeron2001} models, which account for the finite temperature of the WD, yield \EXPN{12.9}{8}, \EXPN{9.1}{8} and \EXPU{6.3}{8}{cm}.  
Both relationships are shown in  Fig.\ \ref{fig_massrad}, and compared to radii
derived from the variable abundance WD plus power law fits.
The average radius (taken from the
time-averaged observations more than 5 days from superoutburst and
excluding Obs.\ 10) for an assumed distance of 65 pc
is \EXPN{10.1}{8}, \EXPN{8.0}{8} and \EXPU{6.3}{8}{cm} for log g=7.5,
8 and 8.5 respectively.\footnote{It is clear from Table \ref{wd80abun}
that the fitted normalizations in the high and low flux phases of
Obs.\ 8 and 9 differ from the normalization measured in the
time-averaged spectra by about 10\%.  This translates to an
uncertainty in the radius of about 5\%. However, the normalizations
derived for the later observations, when there was no time-variable
component agree with those in Obs.\ 8 and 9, when the phase dependent
time variations were evident. } This comparison favors log g=8.5 and a
mass close to 0.93 \MSOL, slightly higher than the values of
8.0$^{+0.38}_{-0.41}$ and 0.71$^{+0.18}_{-0.26}$ \MSOL\ suggested
by \cite{Smith2006}.  However, the results based on our spectral fits depend on the
assumed distance. If VW Hyi is at 74 pc, rather than 65 pc, our
results and those of \cite{Smith2006} can be reconciled.

A distance of 74 pc is certainly within the allowed distance range for VW Hyi, since as we noted earlier,  the current distance estimate is based only on  the relation ship between outburst magnitude and orbital period.  To verify this assertion, we re-estimated the distance to VW Hyi using the latest calibration for the inclination-corrected
M$_{V(max)}$ - P$_{orb}$ relationship for normal dwarf nova outbursts
provided by \cite{Harrison2004}, which agrees very well with the
\cite{Warner1987}  relationship in the period regime appropriate to VW Hyi.  We found
d = 65 $\pm$ 20 pc. This estimate includes the usual inclination correction,
and the quoted uncertainty accounts for both the 0.5 mag scatter in the
relation \cite[see][]{Harrison2004}  and the 10 degree error on the
inclination of VW Hyi. For comparison, a lower limit of 50 pc is obtained
by assuming that the system harbours a non-degenerate secondary which
follows the semi-empirical donor sequence of \cite{Knigge2006} and dominates
the 2MASS K-band flux.

The value derived for \vsini\  and 
F$_{\lambda}$(1000 \AA) for the second component are not dependent on
the adopted log g.  The rotation velocity of the WD, \vsini, appears
to be 
somewhat higher immediately after the outburst, perhaps 500 $\VEL$,
before settling to a value of about 420 $\VEL$ after 5 days.
Similarly, F$_{\lambda}$(1000 \AA), which effectively represents the
residual flux a the base of \LB\, declines rapidly during the first 10
days after the 
outburst and is approximately constant after that time.  It never
disappears entirely, however.  We will return to the question of how
these results should be interpreted in Sec.\ \ref{sec_discussion},
below.

\subsection{The spectrum of the time-variable component -- the Hot
Spot}

It is hard to imagine how the phase-dependent component seen in the
spectrum of VW Hyi following the superoutburst can be associated
directly with the WD, especially since this component peaks at the
same orbital phases where the optical light peaks.  It seems much more
likely that this emission is due to a physically separate source,
associated with the hot spot.  Simple black body fits to the
difference spectra constructed from the high- and low-state spectra in
Obs.\ 8 and 9 yield temperatures of 18,900 K (Obs.\ 8) and 19,600 K (Obs,\
9) and apparent luminosities of \EXPU{9}{31}{\LUM}. This is very
close to the luminosity of the WD at these times. The characteristic
size of the BB emitter inferred from these fits is comparable to, or
somewhat  larger than, that of the WD, i.e. \POW{9}{cm}. While the
exact size depends somewhat on the geometry assumed, as well as on the
isotropy of the emitted radiation, it is clearly much smaller than the
size of the accretion disk ($\simeq$\POW{10}{cm}).

A BB spectrum does not model the observed difference spectra very
well. As indicated in Fig.\ \ref{fig_phase_resolved_spectra}, the
problems are not simply the overall spectral shape but also evidence
for a broad absorption near \LB\ and a narrow feature due
to \CiiiLxi. This suggests modeling in terms of stellar atmospheres.
In order to see if this was an improvement over a BB, we created a
small grid of model spectra with solar abundances and various
gravities ranging from log g=4 to log g=8.  Initial tests showed that
the fits were insensitive to \vsini, mainly because the S/N in the
difference spectra makes it impossible to identify narrow features. We
therefore smoothed the difference spectra to a resolution of 0.5 \AA\
for the purposes of these fits. A comparison of the best fit for log
g=4 is shown in bottom panel of Fig. \ref{fig_second}.  For log g=4,
the best-fit temperature was 25,000 K for both Obs.\ 8 and 9.  The
isotropic luminosity and size scale are smaller than for the blackbody
fits, averaging about \EXPU{2.5}{31}{\LUM} and
\EXPU{3}{8}{cm} for the two observations.  The temperature and size of
the emitting region depend upon the assumed gravity.  For log g=8,
which is less justifiable from a physical perspective than log g=4,
the fitted temperatures are higher, $\sim$40,000K and the sizes are
smaller, \EXPU{~1.1}{8}{cm}. The difference in temperature is an
indication of how much the gravity affects the spectrum of a star in
the FUV in this temperature range.  However, the luminosity is not
very dependent on the assumed gravity, averaging \EXPU{2.5}{31}{\LUM}
for log g=8. The model fits using stellar atmospheres are better fits
than the simple blackbody fits, both qualitatively and in a \CHINU\
sense, but, clearly, from a physical perspective they should still be
viewed with caution.

Nevertheless, if 25,000K is a reasonable estimate for the temperature
of the hot spot emission region observed with \FUSE, it is easy to
show that this emission region does not contribute very much to the
optical emission from the hot spot. More specifically, our 25,000 K
hot spot would produce a flux of only about
\EXPU{7}{-16}{\OIGS} in the visible. By contrast, the amplitude of the
optical modulation (approximately half a magnitude at m$_v$ of
13), corresponds to an optical hot spot flux of \EXPU{9}{-15}{\OIGS}.
Thus the optical hot spot emission must arise from a larger region
emitting with a cooler effective 
temperature. We do not have enough information to accurately measure
the temperature of the optical hotspot, and indeed there have been few
efforts to determine conditions in the physical conditions in the hot
spot.  The existing estimates of hot spot temperatures in other
systems, which are largely based on broad band optical colors, are in the
neighborhood of 10,000 to 15,000 K  \citep[][and references therein]{Warner1995}.  For VW Hyi, these temperatures
and the observed magnitude variation, would imply luminosities for the
optical hot spot of \EXPU{4-9}{31}{\LUM}, comparable to, or slightly
larger than, the luminosity of FUV hot spot.

\subsection{A third component to the spectrum - disk emission}

Orbital variations in the flux from VW Hyi become far less apparent after Obs.\ 9 and
do not reappear after the normal outburst of the system.  However, a
uniform temperature WD does not account for all of the emission from
the system in the FUV, even in Obs.\ 13.  This is illustrated in
Fig.\  \ref{fig_third}, The upper and lower panels of this figure show
the Obs.\ 8 ``low flux'' spectrum and the Obs.\ 13 time-averaged
spectrum with the variable abundance WD model fit subtracted from the
spectrum.  Ignoring the narrow air glow lines, these spectra are
characterized by broad emission features of \CiiiLix, \NiiiL, and a
combination of Lyman$\beta$ and \OviL.  These WD-subtracted spectra
also show a continuum (broken by some absorption features that may
reflect imperfections in the WD models; see discussion above) which
clearly rises towards the Lyman limit, probably due to emission from
the higher-order Lyman lines.  The  
flux associated with this spectrum does not vary greatly between
Obs.\  8 and 13, i.e. from 6 to 30 days after the superoutburst, a
fact which is also evident from the fitted fluxes at 1000 \AA\ in the
variable abundance fits included in Table \ref{wd80abun}.  In the
earlier observations, it is difficult to separate out the continuum
associated with this portion of the spectrum from that associated with
the hot spot, since we do not have good phase coverage.  However, what
is clear from simple inspection of the data is that the flux in \OviL\
is not very different in Obs.\ 2 and 13, so the emission line portion
of the third component is relatively constant over a period from just
after superoutburst and through the normal outburst.

The same third component we have identified here was also apparent in
the \FUSE\ spectrum of VW Hyi that was 
analyzed by \cite{Godon2004}.  Moreover, the same emission lines are
seen in 
some other CVs in quiescence, including SS Cyg and WX Hyi
\citep{Long2005}, and resonance lines of \NvL, \SIivL, and \CivL\ are
commonly seen in the wavelength range covered by \IUE\ and \HST\
\citep{Mauche1997}.  Although the emission lines that are seen in the
optical spectra of SS Cyg and other systems are likely to be
associated with the disk and have been reproduced in simulations of
disk spectra \citep{Williams1980,Kromer2007}, the temperature from a
standard steady state disk is too low to produce emission from lines
like \CivL\ or, especially, \OviL, since the disk is thought to be relatively cold
($<$6000 K) in quiescence.  The broad emission lines seen in the UV
most likely arise in a chromosphere above the surface layer of the
disk.  Fits to the profiles indicate that the resonance lines, such as
\CivL, have surface fluxes in the lines that fall off approximately
as $r^{-2}$ \citep{Long2005}.  The physical mechanism driving a corona
or chromosphere has not been conclusively established.
\cite{Shaviv1986} suggested it could arise from viscous effects in the
upper layers of the disk.  However, \cite{Ko1996} showed that they
could reproduce the line fluxes of a number of systems assuming a
corona generated by X-ray illumination of the disk and favored this
over viscous dissipation effects in the upper atmosphere.  A
re-examination of both of these hypotheses is in order today, given
the existence of a fairly large number of high SN, high resolution
\HST\ spectra and the existence of more detailed theories of the
interaction between the X-ray gas and the inner accretion disk
(e.g. \cite{Meyer2000}). However, the \FUSE\ observations of VW Hyi do
not  provide the ideal set of data to investigate this, since the spectra
are dominated by other spectral components.  Nevertheless, if the FUV
emission lines do arise in the disk, it is quite
likely that the rest of the third component spectrum can be explained
away as emission from weaker lines, plus possibly some continuum
emission from the disk corona.  There is no compelling reason to
associate it with the boundary layer, if by the boundary layer one means a
localized structure at the interface between the accretion disk and
the WD.

\section{Discussion\label{sec_discussion}}

When the disks in dwarf novae transition into the outburst state, the
WD is buffeted by the effects of radiation and increased mass transfer
onto the WD surface.  In the case of VW Hyi, this buffeting lasted
about 9 days and heated the WD by at least 5,000 K.  The WD was far
less affected by the much shorter normal outburst that occurred prior
to Obs.\ 12.  Similarly, the outburst appears to have buffeted the
secondary as well, as evidenced by the fact that the hot spot was
considerably brighter after the outburst than previously, resulting in
a hotter and more extensive hot spot that even affected the FUV
emission.  The effects of the normal outburst on the secondary were
clearly much weaker.

The temperature evolution of the WD in VW Hyi was shown in Fig.\
\ref{fig_trends}.  The luminosity of the WD component in the spectrum
is shown in Fig.\ \ref{fig_wdlum}.  Just after the outburst, the
luminosity of the WD was about \EXPU{1.7}{32}{\LUM}. It then declined by
almost a factor of 3 by the time of the last observation with \FUSE.
The decay was approximately exponential, with a time constant of about 7
days.  If L$_{wd}$ at Obs.\ 2 is interpreted as the quiescent
luminosity, then the luminosity excess was \EXPU{1.1}{32}{\LUM}.  These
results are very similar to the results obtained by
\cite{Gaensicke1996} based on a collection of \IUE\ observations of VW
Hyi (even though the \IUE\ data did not allow a clean separation of
the WD from the other components in the emission spectrum).
\cite{Gaensicke1996} found a higher temperature of 26,400 K than we
did here just after outburst, and this may be due to the effects of
other continuum components.  Had we neglected the other components in
our analysis, we would have derived temperatures that were higher by
several thousand degrees just after superoutburst.  But it could also
be that the superoutburst we observed was somewhat shorter than the
typical superoutburst.  Either of these effects could also explain
why the time constant for WD cooling we derive -- 7 days -- was somewhat
shorter than the value they derived, which was 9.8$^{+1.4}_{-1.1}$ days.

Three closely related explanations have been suggested for the
temperature evolution of a WD in a CV in the aftermath of an
outburst. The first concentrates on the altered state of the WD after
the outburst.  More 
specifically, the outburst heats the surface layers of the WD, and,
once the outburst ceases, the WD radiates away the thermal energy that
was deposited. The outburst also deposits material on the
surface. Even if this material were deposited 
on the photosphere at the photospheric temperature, it would still
leave the WD out of equilibrium structurally.  Once the outburst
ceases, the surface 
layers are able to radiate the thermal energy that was deposited and, 
over longer time scales, the WD will readjust its internal structure
to reflect its increase in mass \citep{Sion1995, Piro2005}.  The fact
that WDs in CVs accrete and grow in mass (at least between nova
outbursts) explains why they are hotter than isolated WDs of
comparable age \citep{Townsley2003}.

The second explanation for WD heating was advanced by \cite{Long1993}, 
who suggested that departures from single temperature WD fits to the
HUT spectrum of U Gem in quiescence could be explained if the outburst
spun-up a surface layer of the WD and the slow conversion of
rotational energy to heat produced a gradually decaying apparent
temperature for the WD.  Two component WD fits to the quiescent
spectra of several DNe, including VW Hyi, have been carried out, and
do produce improvements in the statistical quality of some of the
fits.  Generally speaking, as in the case of VW Hyi \citep{Sion2001},
the second component appears to be stronger just after an outburst and
to rotate more rapidly than the primary component. One of the
arguments for this hypothesis has been that it helps to explain
away a troubling problem for single temperature fits in U Gem, which
was that the radius 
of the WD appeared to evolve with time from outburst.  However, despite the
fact that two-component WD fits produce lower \CHINU, this is not
strong evidence for the physical reality of a rotating region of the
WD, especially given our increased awareness of multiple components in the FUV spectra of quiescent CVs.  For example, it is easy to demonstrate the same improvement in
\CHINU\ can be achieved with a model consisting of the WD and a power law. To
assert physical reality, one needs clear evidence of more rapid
rotation in the higher ionization state lines or in the FUV portion of
the spectrum.  Although VW Hyi clearly has multiple components in its
spectra, there is no ``smoking gun'' for a second component with a
spectrum resembling a rapidly rotating, hotter region of the WD, or
indeed any other component, such an an inner accretion disk ring, with
a different temperature from the WD.  Furthermore, \cite{Piro2005} has
recently investigated this possibility that rapidly rotating belts
exist in the WD surface layers, concluding that although they may
exist, they are unlikely to be as important as compression heating in
explaining the cooling time of the WD.

The third scenario for explaining the time evolution of the WD 
spectrum posits that the accretion rate is still elevated at the end
of the optical outburst and that the $T_{wd}$ is elevated by this
accretion.  There is some theoretical basis for this. In particular,
\cite{Meyer2000}, partly to explain the fact that X-ray outbursts of
DNe are often delayed from the optical outbursts, proposed that
accretion onto the WD in quiescence proceeds via a coronal flow from
the disk onto the WD, and that the coronal flow erodes the inner
portions of the quiescent disk, leaving a hole in the inner disk.  As described
by \cite{Mineshige1998}, the hole steadily expands toward larger radii in quiescence
and the mass accretion rate (see their equation 4) onto the WD drops with time.
If this is the case, then both the X-ray measured accretion rate and possibly the
WD temperature should drop with time.   Observationally, \cite{vanderWoerd1987} observed
VW Hyi for an extended period in 1986 and concluded the X-ray flux
declined slowly by a factor of 1.2-1.6 during the inner outburst
period, which they linked to a decline in the mass accretion
rate.  \cite{Pandel2003} used XMM to observe VW Hyi in mid-quiescence
and showed that the X-ray emission lines which dominated the X-ray spectrum may
be rotating at the velocity of the WD, but not at the velocity of the
inner accretion disk.  They also found that short term variations in the X-ray flux lag short term variations in the 2400-3400 \AA\ flux by about 100 seconds.  They modeled the emission in terms of a cooling
flow onto the WD surface from the disk, all of which is generally consistent with the
picture proposed by \cite{Meyer2000}.  \cite{Liu2008} have recently
modeled the X-ray spectrum of VW Hyi quiescence in terms a hot
tenuous coronal inflow, concluding that one can obtain good fits to
the data, but only if the conductivity of the plasma is low.

To our knowledge, no one has actually attempted to calculate the
heating of the WD due to a coronal flow.  That said, it seems unlikely
that ongoing accretion in VW Hyi plays the dominant role in the
temperature evolution of the WD, mainly because the required accretion
rate is so substantial. There is very little evidence for such a
large accretion rate from the inner disk in quiescence, or just after the return to
quiescence.  Based on STIS spectra of VW Hyi, \cite{Merritt2007}
estimate the accretion rate in the inner disk to be
\EXPU{\sim4}{-9}{\MSOL yr^{-1}}, assuming $M_{wd}$ of 0.8 \MSOL and a
distance of 65 pc.  We plan, as noted earlier, a full discussion of
the \FUSE\ outburst spectra later.  However, following a procedure
essentially identical to that of \cite{Merritt2007}, we derive a mass
accretion rate $\dot{m}_{disk}$ during Obs.\ 1 of about
\EXPU{0.9}{-9}{\MSOL \: yr^{-1}}, which corresponds to a disk
luminosity of \EXPU{3.4}{33}{\LUM}. Our accretion rate is lower because the flux from VW Hyi was lower than in the observations described by \cite{Merritt2007}.  If one wished to attribute the
excess luminosity of the WD during Obs.\ 2 to instantaneous heating of
the WD by the ongoing accretion, one would need the accretion rate to
be at least 0.75\% of that observed at outburst maximum, or 3\% of the
rate when measured during Obs.\ 1, that is \EXPU{2.7}{-11}{\MSOL \:
yr^{-1}}.  .
This would have to be mostly hidden, since the accretion
rate at the inner boundary reported by \cite{Pandel2003} from XMM at
mid-quiescence was far less, namely \EXPU{8}{-14}{\MSOL yr^{-1}}.  It is interesting that the overall abundances that we measure in VW Hyi
are higher than in some other dwarf novae, which might indicate the accretion rate
is higher than normal. But the absence of significant and continuing decline in the 
abundances with time from outburst suggests that the accretion rate is not
varying dramatically.  In principle, given an accretion rate and the abundances of the accreting material, one should be able to calculate the photospheric abundances, but this calculation has also not, to our knowledge, been carried out.

It is interesting in this regard that the time scale associated with declining emission from the hot spot observed optically and with FUSE in VW Hyi is similar to the WD cooling time scale.  The mass  transfer rate from the secondary to the hotspot, given a luminosity of \EXPU{1}{32}{\LUM } is about \EXPU{2}{-9}{\MSOL \:
yr^{-1}} \cite[see, e. g.][to obtain this estimate]{Warner1995}.  Thus. if a portion of this material could make its way to the WD, it would provide a source of heating.\footnote{As we noted earlier the luminosity of the hot spot and of the WD are similar.  This, however, is completely coincidental.  If the accretion rate onto the WD were similar to that of hot spot, then the WD would have a temperature of order 60,000K, which is ruled out by the observations.}  The Doppler maps of the VW Hyi in quiescence obtained by \cite{Smith2006} do show that  emission from the stream is extended in the region between the stream trajectory and its so-called ``Keplerian shadow'' of the stream, but do not appear to show heated disk material at the WD.  Hydrodynamic simulations of stream disk interactions \citep[e. g.][]{Armitage1996} show material richocheting off of the stream impact region to significant heights above the disk, and extending toward smaller radii.  But, ultimately, this material is entrained in the disk, and does not end up on the WD.   As in the case of the coronal flow picture developed by \cite{Meyer2000} and recently described by \cite{Liu2008}, it is not clear how this material could remain undetected in X-rays. Our view is that it will be very difficult to establish that strong, ongoing accretion onto the WD is occuring in VW Hyi through studies of WD cooling, since the short interoutburst period limits the baseline for such studies. However, \cite{Godon2006}  have argued that ongoing accretion {\em is} required to explain the cooling of the WD in WZ Sge, which exhibits far more infrequent outbursts than VW Hyi.

There have also been a number of suggestions that the boundary layer
might be present in the UV spectra of DNe, including VW Hyi.
\cite{Godon2005} present a recent summary of the case for this in VW
Hyi. The point out that a number of studies with \HST\ and \FUSE\
have concluded that a non-WD component contributes to the quiescent spectrum
of VW Hyi. They also note that when this component is modelled with WD
spectra, it typically yields a higher temperature and faster rotation
speed than the actual WD component. \cite{Godon2004}, in particular,
analyzed a spectrum obtained with of VW Hyi obtained 11 days after a normal outburst and concluded that the non-WD
component had a temperature of 50,000 K and \vsini\ of 3000 $\VEL$, similar
to the Keplerian velocity in the inner accretion disk.  We agree that
there is a very evident non-WD component in the spectrum analyzed by
\cite{Godon2004} (indeed this was one of the motivations for
carrying out the present study). 
We have re-examined this spectrum, and checked, in particular, whether
there was any evidence of orbital modulation; there was
not. Interestingly, the spectral shape of the non-WD component in this
spectrum seems to resemble the spectral shape of what we have called the third
component in our spectra, i.e. relatively flat with a rise in flux towards the Lyman
limit. However, this component is much brighter in the
\cite{Godon2004} spectrum than in our spectra following the normal
outburst (e.g. Obs.\ 13). \cite{Godon2005} assert that it seems likely
to them that this component is the boundary layer and that this
component provides 4/5 of the boundary layer luminosity. 

Our view, on the other hand, is that the data do not constrain the
origin of this component of the spectrum, because, except for the
emission lines, which arise from a substantial portion of the disk, the
spectrum of what we have called the ``third component'' is relatively
featureless. In order to establish that the continuum arises from the boundary
layer, one needs to show observationally that it is not associated with
the disk, or at least show that the emission lines can be produced
without producing an associated continuum, or pseudo-continuum of
other lines.

Finally, it is clear to us that a precise definition of what one means
by boundary layer emission is sorely needed.  This definition needs to
distinguish direct emission from the boundary layer from
various forms 
of reprocessed emission, including a coronal surface layer on the
surface layer of the disk.  In the absence of such a definition, it
will be difficult for observers and theorists to reach consensus on
the nature of the multiple components in the spectrum of VW Hyi and
other systems in quiescence.

\section{Conclusions}

We have carried a detailed study of the SU UMa system VW HYi 
using \FUSE. This is most intensive FUV study ever undertaken with a
view to understanding the
effect of an outburst on the WD in a nonmagnetic cataclysmic
variable. In particular, we have observed the system from the end of a
superoutburst all the way beyond the next normal outburst. Our main
conclusions are as follows:

\begin{itemize}

\item The quiescent spectra are dominated by the WD, as anticipated,
but also contain additional components. The WD is heated by the
superoutburst to a temperature of 24,000 K and decays back to
quiescence with a time constant of approximately 8 days; this confirms
confirming the conclusions of \cite{Gaensicke1996}. The WD was not heated
significantly by the normal outburst.  It is rotating with \vsini
of 420 $\VEL$, and our abundance analysis confirms that the
photosphere is deficient in C relative to N, again confirming earlier
results obtained with \HST\  that CNO processed material is being accreted 
onto the WD. The inferred abundance variations as
a function of time from outburst are small, provided one allows for 
a second component in the FUV flux when fitting the data.
\item There are phase-dependent
FUV flux variations following the superoutburst, which last
for at least 10 days beyond the end of the superoutburst.   The strongest modulation 3 to 5 days after return to quiescence occurs on the superhump period, but later on the power in the
orbital period is dominant, mimicking the behavior of the system at optical wavelengths during this period.  The FUV flux peaks near 
phase 0.8, at the same phase where the hot spot is brightest at
optical wavelengths; it therefore seems likely that the variable FUV
flux component is associated with the hot spot.  The FUV spectrum
of the hot spot is relatively featureless and
significantly hotter than either the optical hot spot or the WD in VW
Hyi.
\item The spectrum of VW Hyi also contains a third component in
the FUV, which is seen clearly in the broadened emission lines
associated with the disk.  This third component appears during all
phases of the quiescent observations and appears to have an
associated "pseudo-continuum", which may be composed of large number
of weaker lines or a true continuum.  There is no compelling evidence
for significant FUV emission from the boundary layer.  
\end{itemize}

\acknowledgments{This analysis of {\it FUSE} data would not have been
possible without the financial support from NASA through grant
NNG04GQ38G. We appreciate this, as well as the dedicated efforts of
the entire {\it FUSE} team who spent the time and effort to make sure
that the observations were a success.  Janet A. Mattei of the AAVSO notified us that 
VW Hyi had undergone superoutburst and counseled us on when to schedule the observations; though others carry on admirably at the AAVSO, we miss her expertise, dedication, and friendship.}


\clearpage

\begin{deluxetable}{rcccrrr}
\tablecaption{Observation Log }
\tablehead{\colhead{Obs.~\#} & 
 \colhead{Dataset} & 
 \colhead{Start~Date} & 
 \colhead{Start~Time} & 
 \colhead{Time~from~T$_o^a$} & 
 \colhead{Side~1~Exp.} & 
 \colhead{Side~2~Exp} 
\\
\colhead{~} & 
 \colhead{(UT)} & 
 \colhead{(UT)} & 
 \colhead{(days)} & 
 \colhead{(s)} & 
 \colhead{(s)} & 
 \colhead{} 
}
\tabletypesize{\scriptsize}
\tablewidth{0pt}\startdata
1 &  E1140101 &  2004-07-30 &  14:52:02 &  -0.88 &  5939 &  5936\tablenotemark{b} \\ 
2 &  E1140102 &  2004-07-31 &  12:22:42 &  0.02 &  6320 &  6320 \\ 
3 &  E1140103 &  2004-08-01 &  19:51:48 &  1.33 &  3212 &  3266 \\ 
4 &  E1140104 &  2004-08-02 &  10:28:01 &  1.94 &  5558 &  5603 \\ 
5 &  E1140105 &  2004-08-03 &  09:26:24 &  2.89 &  4809 &  4818 \\ 
6 &  E1140106 &  2004-08-04 &  06:16:17 &  3.76 &  8378 &  8357 \\ 
7 &  E1140107 &  2004-08-05 &  07:11:06 &  4.80 &  1100 &  6083 \\ 
8 &  E1140108 &  2004-08-07 &  00:40:53 &  6.53 &  22754 &  22733 \\ 
9 &  E1140109 &  2004-08-10 &  03:19:43 &  9.64 &  24018 &  24065 \\ 
10 &  E1140110 &  2004-08-13 &  10:40:43 &  12.94 &  27160 &  26966 \\ 
11 &  E1140111 &  2004-08-16 &  06:28:24 &  15.77 &  19496 &  19278\tablenotemark{b} \\ 
12 &  E1140112 &  2004-08-19 &  05:41:46 &  18.74 &  27254 &  26720 \\ 
13 &  E1140113 &  2004-08-30 &  01:08:05 &  29.55 &  21916 &  21894 \\ 
\tablenotetext{a}{ T$_o$ is time from JD2453218.0, see text.}
\tablenotetext{b}{Outburst spectrum.}
\enddata 
\label{ObsLog}
\end{deluxetable}

\begin{deluxetable}{rccrrrrl}
\tablecaption{Scaled Abundance WD - Power Law Fits }
\tablehead{\colhead{Obs.~\#} & 
 \colhead{Norm.} & 
 \colhead{$T_{wd}$} & 
 \colhead{z} & 
 \colhead{v~sin(i)} & 
 \colhead{f(1000~\AA)$^a$} & 
 \colhead{$\alpha$} & 
 \colhead{$\CHINU$} 
\\
\colhead{~} & 
 \colhead{($10^{-23}$)} & 
 \colhead{(1000~K)} & 
 \colhead{~} & 
 \colhead{($\VEL$)} & 
 \colhead{~} & 
 \colhead{~} & 
 \colhead{~} 
}
\tabletypesize{\scriptsize}
\tablewidth{0pt}\startdata
2 &  21.9$\pm$1.9 &  24.3$\pm$0.2 &  3.3$\pm$1.0 &  530$\pm$170 &  10.3$\pm$0.4 &  2.2$\pm$1.0 &  3.6 \\ 
3 &  21.9$\pm$2.2 &  23.3$\pm$0.3 &  2.1$\pm$0.6 &  430$\pm$190 &  3.8$\pm$0.2 &  3.1$\pm$1.4 &  2.1 \\ 
4 &  22.1$\pm$1.6 &  23.1$\pm$0.2 &  2.6$\pm$0.8 &  450$\pm$150 &  5.1$\pm$0.2 &  3.3$\pm$1.0 &  2.9 \\ 
5 &  21.4$\pm$2.0 &  22.4$\pm$0.2 &  1.9$\pm$0.6 &  420$\pm$240 &  3.2$\pm$0.2 &  2.5$\pm$1.3 &  2.4 \\ 
6 &  20.7$\pm$2.2 &  22.3$\pm$0.2 &  1.8$\pm$0.6 &  360$\pm$220 &  3.2$\pm$0.2 &  2.5$\pm$1.4 &  3.0 \\ 
7 &  19.5$\pm$1.9 &  22.0$\pm$0.2 &  2.2$\pm$0.8 &  450$\pm$190 &  2.5$\pm$0.2 &  3.8$\pm$1.6 &  1.7 \\ 
8 &  18.4$\pm$2.5 &  22.0$\pm$0.3 &  1.8$\pm$0.7 &  360$\pm$230 &  3.3$\pm$0.1 &  3.1$\pm$1.0 &  5.8 \\ 
8~(high) &  16.9$\pm$2.5 &  22.6$\pm$0.3 &  2.1$\pm$0.7 &  430$\pm$200 &  5.0$\pm$0.2 &  4.1$\pm$0.8 &  2.6 \\ 
8~(low) &  19.7$\pm$1.7 &  21.4$\pm$0.2 &  1.6$\pm$0.5 &  350$\pm$200 &  1.1$\pm$0.1 &  6.0$\pm$2.3 &  3.1 \\ 
9 &  17.8$\pm$2.1 &  21.2$\pm$0.2 &  1.4$\pm$0.5 &  360$\pm$230 &  2.1$\pm$0.1 &  3.1$\pm$1.2 &  5.5 \\ 
9~(high) &  15.7$\pm$1.6 &  22.0$\pm$0.3 &  1.5$\pm$0.4 &  360$\pm$160 &  3.7$\pm$0.2 &  3.1$\pm$0.8 &  1.9 \\ 
9~(low) &  18.6$\pm$2.3 &  20.9$\pm$0.2 &  1.4$\pm$0.5 &  350$\pm$260 &  0.8$\pm$0.1 &  5.5$\pm$2.1 &  3.6 \\ 
10 &  17.0$\pm$1.9 &  20.8$\pm$0.2 &  1.3$\pm$0.5 &  350$\pm$190 &  1.2$\pm$0.1 &  3.6$\pm$1.9 &  5.6 \\ 
12 &  18.1$\pm$2.6 &  20.6$\pm$0.3 &  1.3$\pm$0.6 &  360$\pm$240 &  1.7$\pm$0.1 &  2.5$\pm$1.6 &  6.1 \\ 
13 &  15.4$\pm$2.2 &  19.5$\pm$0.2 &  0.8$\pm$0.5 &  320$\pm$280 &  0.9$\pm$0.1 &  2.1$\pm$2.1 &  4.2 \\ 
\tablenotetext{a}{ f(1000) \AA\ is in units of 10$^{-14}$ $\OIGS$}
\enddata 
\label{wd80pow}
\end{deluxetable}

\begin{deluxetable}{rccrrrrrrrl}
\tablecaption{Variable Abundance WD - Power Law Fits }
\tablehead{\colhead{.~\#} & 
 \colhead{Norm.} & 
 \colhead{$T_{wd}$} & 
 \colhead{C} & 
 \colhead{N} & 
 \colhead{Si} & 
 \colhead{S} & 
 \colhead{v~sin(i)} & 
 \colhead{f(1000~\AA)$^a$} & 
 \colhead{$\alpha$} & 
 \colhead{$\CHINU$} 
\\
\colhead{~} & 
 \colhead{($10^{-23}$)} & 
 \colhead{(1000~K)} & 
 \colhead{~} & 
 \colhead{~} & 
 \colhead{~} & 
 \colhead{~} & 
 \colhead{($\VEL$)} & 
 \colhead{~} & 
 \colhead{~} & 
 \colhead{~} 
}
\tabletypesize{\scriptsize}
\tablewidth{0pt}\startdata
2 &  21.6$\pm$1.4 &  24.1$\pm$0.2 &  0.8$\pm$0.2 &  6.4$\pm$2.2 &  4.4$\pm$1.0 &  2.8$\pm$0.5 &  570$\pm$30 &  10.4$\pm$0.3 &  1.5$\pm$0.7 &  3.2 \\ 
3 &  22.4$\pm$1.3 &  23.0$\pm$0.2 &  0.6$\pm$0.2 &  5.7$\pm$2.2 &  2.8$\pm$0.5 &  1.3$\pm$0.5 &  460$\pm$30 &  3.9$\pm$0.2 &  2.5$\pm$1.1 &  1.9 \\ 
4 &  22.7$\pm$1.4 &  22.8$\pm$0.2 &  0.6$\pm$0.2 &  6.9$\pm$2.2 &  3.8$\pm$0.8 &  1.8$\pm$0.6 &  490$\pm$30 &  5.2$\pm$0.3 &  2.5$\pm$0.9 &  2.5 \\ 
5 &  22.7$\pm$1.2 &  22.1$\pm$0.2 &  0.5$\pm$0.2 &  5.8$\pm$1.7 &  3.4$\pm$0.8 &  1.2$\pm$0.4 &  450$\pm$20 &  3.4$\pm$0.2 &  2.2$\pm$1.1 &  2.0 \\ 
6 &  21.6$\pm$1.2 &  22.0$\pm$0.2 &  0.5$\pm$0.2 &  4.6$\pm$1.9 &  2.8$\pm$0.5 &  1.1$\pm$0.3 &  410$\pm$20 &  3.3$\pm$0.1 &  2.2$\pm$0.9 &  2.5 \\ 
7 &  20.4$\pm$1.4 &  21.6$\pm$0.2 &  0.5$\pm$0.3 &  4.4$\pm$1.6 &  3.4$\pm$0.9 &  2.2$\pm$0.7 &  490$\pm$30 &  2.6$\pm$0.1 &  3.1$\pm$0.9 &  1.5 \\ 
8 &  19.7$\pm$1.3 &  21.6$\pm$0.2 &  0.5$\pm$0.2 &  4.0$\pm$2.1 &  2.7$\pm$0.4 &  1.0$\pm$0.3 &  410$\pm$30 &  3.4$\pm$0.1 &  2.2$\pm$0.9 &  4.8 \\ 
8~(high) &  17.6$\pm$1.3 &  22.3$\pm$0.3 &  0.7$\pm$0.2 &  5.8$\pm$2.4 &  2.7$\pm$0.7 &  1.3$\pm$0.4 &  470$\pm$40 &  5.0$\pm$0.2 &  3.4$\pm$0.6 &  2.5 \\ 
8~(low) &  21.5$\pm$1.0 &  21.1$\pm$0.1 &  0.5$\pm$0.2 &  4.8$\pm$1.7 &  2.8$\pm$0.4 &  1.0$\pm$0.2 &  400$\pm$20 &  1.2$\pm$0.1 &  4.3$\pm$1.5 &  2.4 \\ 
9 &  19.9$\pm$1.2 &  20.8$\pm$0.2 &  0.4$\pm$0.2 &  4.9$\pm$2.0 &  2.7$\pm$0.4 &  1.0$\pm$0.2 &  410$\pm$30 &  2.2$\pm$0.1 &  2.5$\pm$0.9 &  4.0 \\ 
9~(high) &  18.0$\pm$1.3 &  21.5$\pm$0.2 &  0.6$\pm$0.2 &  6.6$\pm$2.2 &  2.4$\pm$0.4 &  1.0$\pm$0.3 &  440$\pm$40 &  3.8$\pm$0.2 &  2.5$\pm$0.8 &  1.7 \\ 
9~(low) &  21.5$\pm$1.1 &  20.4$\pm$0.2 &  0.4$\pm$0.2 &  5.4$\pm$1.9 &  2.8$\pm$0.5 &  1.0$\pm$0.1 &  410$\pm$30 &  0.9$\pm$0.1 &  4.6$\pm$1.4 &  2.5 \\ 
10 &  20.1$\pm$1.1 &  20.3$\pm$0.1 &  0.5$\pm$0.2 &  6.0$\pm$2.2 &  2.7$\pm$0.4 &  1.0$\pm$0.2 &  420$\pm$40 &  1.3$\pm$0.1 &  2.5$\pm$1.3 &  4.0 \\ 
12 &  21.0$\pm$1.3 &  20.2$\pm$0.2 &  0.4$\pm$0.2 &  5.8$\pm$2.2 &  2.7$\pm$0.3 &  1.0$\pm$0.3 &  440$\pm$30 &  1.8$\pm$0.1 &  2.2$\pm$1.0 &  4.5 \\ 
13 &  19.4$\pm$1.3 &  19.0$\pm$0.2 &  0.4$\pm$0.2 &  6.1$\pm$2.2 &  2.5$\pm$0.6 &  1.0$\pm$0.1 &  410$\pm$30 &  0.9$\pm$0.1 &  2.1$\pm$1.3 &  3.2 \\ 
\tablenotetext{a}{ f(1000) \AA\ is in units of 10$^{-14}$ $\OIGS$}
\enddata 
\label{wd80abun}
\end{deluxetable}

\pagebreak

\clearpage

\figcaption[aavso]{Optical and ultraviolet light curves of VW\,Hyi obtained
throughout August 2004. Time-resolved ground-based observations are
shown in black (a--n), \FUSE\ observations in blue (01--13), and
observations from the AAVSO data base in red (filled circles are
detections, open circles are upper limits). \label{fig_aavso}}

\figcaption[spectra]{
The time-averaged spectra obtained during Obs.\ 1, 2, 8, and 13.  All of the spectra have been scaled so that the flux at 1100 \AA\ is at about the same relative position each frame.  The flux observed from VW Hyi declined steadily after the superoutburst, until Obs.\ 11 when the source was observed near the end of a normal outburst.  The Obs.\ 11 spectrum was essentially identical to that observed in Obs.\ 1.  By Obs. 13, the flux had declined to a level below that observed in Obs.\ 10.  The character of the spectrum obtained at the end of the super-outburst is quite different from the spectra observed when the system was in optical quiescence.  It has a narrower \LB\  profile and far fewer deep absorption features in the wavelength range between 1050 and 1160 \AA.  The outburst spectra show \OviL\  in absorption whereas the quiescent spectra all show it in emission. All of the spectra, including those in quiescence, show emission extending to the Lyman limit, although the relative intensity of emission at the shortest wavelengths declines with time from outburst. The narrow emission lines which appear in the spectra are all airglow lines.  
\label{fig_spectra}
}

\figcaption[tsa_berto]{Time-series analysis of the optical
data, power spectra on the left, and phase-folded light curves on the
right.  $P_\mathrm{orb}$ and $P_\mathrm{sh}$ indicate the orbital and
superhump period, respectively. From top to bottom: (a) analysis of
combined optical observations b--h (see Fig.\,\ref{fig_aavso}). The
light curve is folded on the period corresponding to maximum power, the
zero-point in phase is arbitrary. (b) same as top panel, but after
removing the superhump signal. The light curve is folded on the
ephemeris in Eq.\,(\ref{e-eph}).  (b) same as top panel, but
after removing the orbital signal. The light curve is folded on the
superhump period, the zero-point in phase is arbitrary. (d) power
spectrum and phase-folded light curve for the combined optical
observations j--n, the light curve is folded on the ephemeris
of \citet{VanAmerongen1987}.  \label{tsa_berto} }

\figcaption[tsa_fuse]{Time-series analysis of the \FUSE\
data. $P_\mathrm{orb}$ and $P_\mathrm{sh}$ indicate the orbital and
superhump period, respectively. (a) analysis of the combined
observations 03--05 (see Fig.\,\ref{fig_aavso}). The light curve is
folded on the superhump period,  the zero-point in phase is arbitrary. (b)
analysis of the combined observations 08--10, the light curve is
folded on the ephemeris in Eq.\,(\ref{e-eph}). \label{tsa_fuse} }

\figcaption[lightcurve]{
 The average flux from VW Hyi measured as a function of time in Obs.\ 8 and 9. The flux is obtained from spectra extracted in 1 minute intervals, ignoring regions where airglow is likely to appear.  Time is measured in periods from a fiducial time prior to the first observation of VW Hyi.  Integral periods correspond to secondary conjunction, which occurs 0.15 periods after phase 0 in the ephemeris of \cite{VanAmerongen1987}.   The gaps are times without data, often due to earth blockage. The observed flux is preferentially high just prior to secondary conjunction, near phase 0.8, that is near hot spot maximum  \label{fig_lightcurve} }

 \figcaption[phase_resolved]{ Spectra of VW Hyi from Obs.\ 8 and 9 in
 black and red, respectively.  The upper panel shows the ``high flux''
 spectrum during orbital phase 0.625-1.0 when the overall flux was
 highest.  The middle panel shows the ``low flux'' spectrum from
 phases 0.0 to 0.375 when the flux was low.  The lower panel shows the
 difference spectrum. Note that the vertical scale for the panel
 showing the difference spectrum is not the same as for the other two
 panels.  All of the spectra have been box-car smoothed to a
 resolution of about 0.5 \AA.  \label{fig_phase_resolved_spectra} }

\figcaption[vwhyi08_z_abun] {A comparison of two model fits to the time-averaged spectrum of Obs.\ 8.  In both cases  a log g=8 WD - models were fit to the data.  In the upper panel, only the overall metal abundance was varied while in the lower panel the individual abundances of C, N, Si and S were varied, and all other species were set to 1.     The data are plotted in black if this portion of the data was used in the fit and in grey if the data were excluded from the fit.  The best fitting model is plotted in blue.   The power-law contribution to the total spectrum is shown in red in the lower panel. \label{fig_vwhyi8_metal_var}
}

\figcaption[vwhyi_abun]{Log g=8 WD - power law fits to the time-averaged spectra from Obs.\ 2 and 13.  In both cases, the abundances of C, N, Si and S were allowed to vary independently. 
The color conventions are the same as in Fig.\  \ref{fig_vwhyi8_metal_var}.\label{fig_vwhyi_abun}}

\figcaption[vwhyi8_phase_abun]{Log g=8 WD - power law fits to the high (upper panel) and low (lower panel) flux spectra obtained of VW Hyi in Obs.\ 8.  The abundances are allowed to vary in both cases.  The color conventions are the same as in Fig.\  \ref{fig_vwhyi8_metal_var}. \label{fig_vwhyi8_phase_abun}
}

\figcaption[vwhi_sum]{A summary of trends in T$_{wd}$ (upper left), \vsini\  (upper right), R$_{wd}$ (lower left), and the F$_{\lambda}$(1000 \AA) flux from the second component  (lower right) as a function of time from superoutburst when fitting the data as a WD with variable abundances and a power law.  The black, blue and red curves curves correspond to log g=7.5, 8.0, and 8.5 models respectively.  The dotted horizontal lines in the lower left hand panel correspond from top to bottom the expected radius for a log g=7.5, 8.0 and 8.5 WD, based on the models of \cite{Bergeron2001} for a 20,000K WD. \label{fig_trends}
}

\figcaption[vwhyi_massrad]{The radius of the WD in VW Hyi as measured from fits to log g=7.5, 8.0 and 8.5 variable abundance models assuming a distance of 65 pc.  Only fits after day 5 are shown.  Mass radius relations for a cold \citep{hamada1961} and 20,000K \citep{Bergeron2001} WD are shown as the dashed black and solid red curves, respectively.  The comparison suggests a somewhat higher mass or somewhat greater distance for VW Hyi than is commonly assumed. \label{fig_massrad}
}

\figcaption[vwhyi_second]{Comparisons of the best fitting BB (upper panel) and log g=4 stellar models to the ``hotspot'' spectrum as measured in Obs.\  8. 
\label{fig_second}}

\figcaption[vwhyi_third]{A comparisons between the non-WD components of the Obs.\ 8 ``low flux'' (top panel) and Obs.\ 13 time-averaged (bottom panel) spectra to the ``hot spot'' spectrum obtained in Obs.\  8.  All of the spectra have been box-car smoothed to a resolution of about 1 \AA. The very narrow emission features in the upper and low panel are all air-glow lines, and the same lines occasionally manifest themselves as narrow absorption profiles in the hot spot spectrum. \label{fig_third}}

\figcaption[wd_lum]{ The luminosity (plotted in red) of the WD as a function of time from outburst as calculated from the  log g=8 variable abundance plus power law fits.  The  black lines are for an models with exponential decays of 5, 7 and 9 days.  The data point which falls above all 3 exponential curves was for the observation that occurred just after the normal outburst of VW Hyi \label{fig_wdlum}}

\newpage
\pagestyle{empty}

\clearpage

\plotone{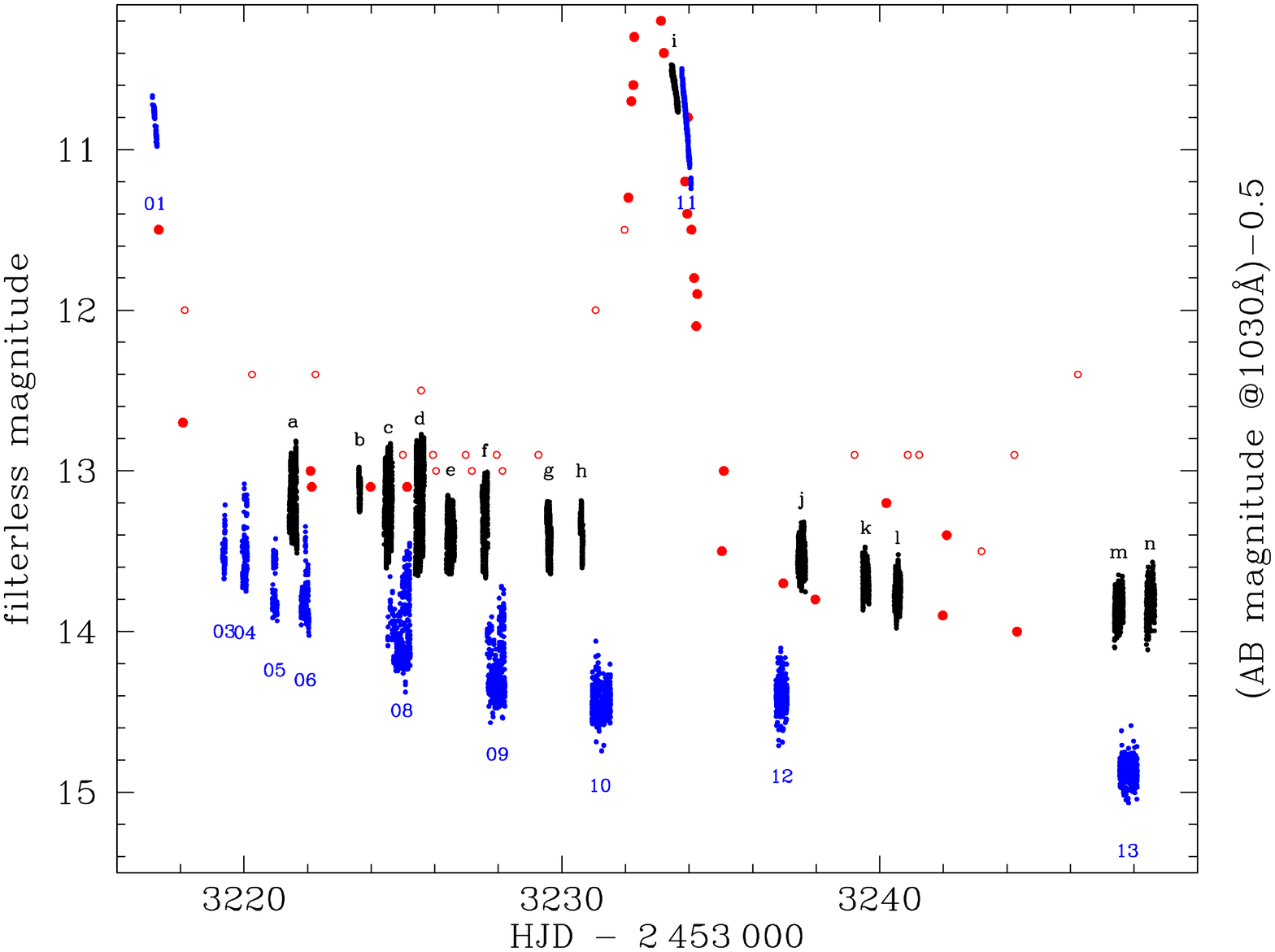}

\plotone{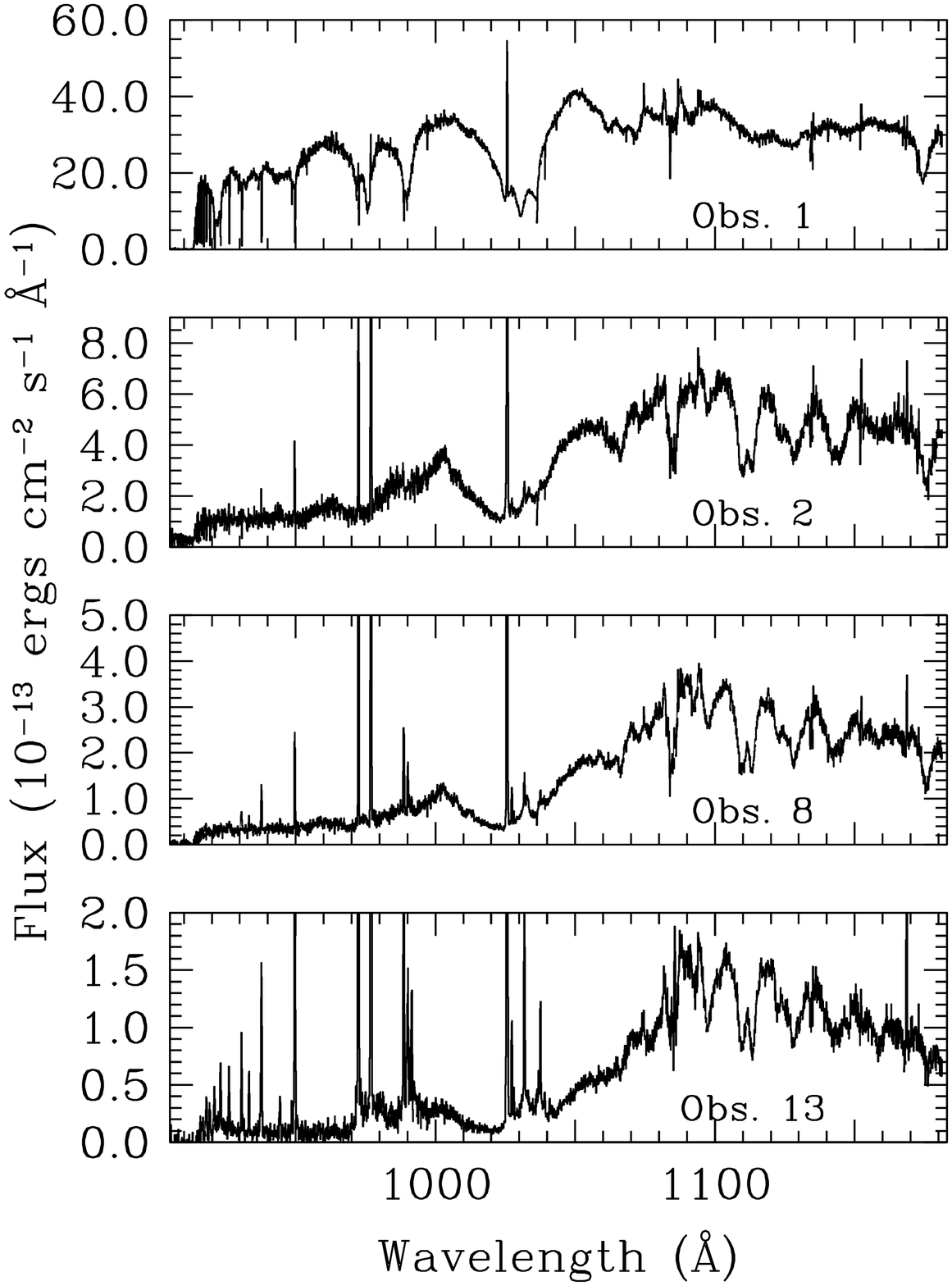}

\plotone{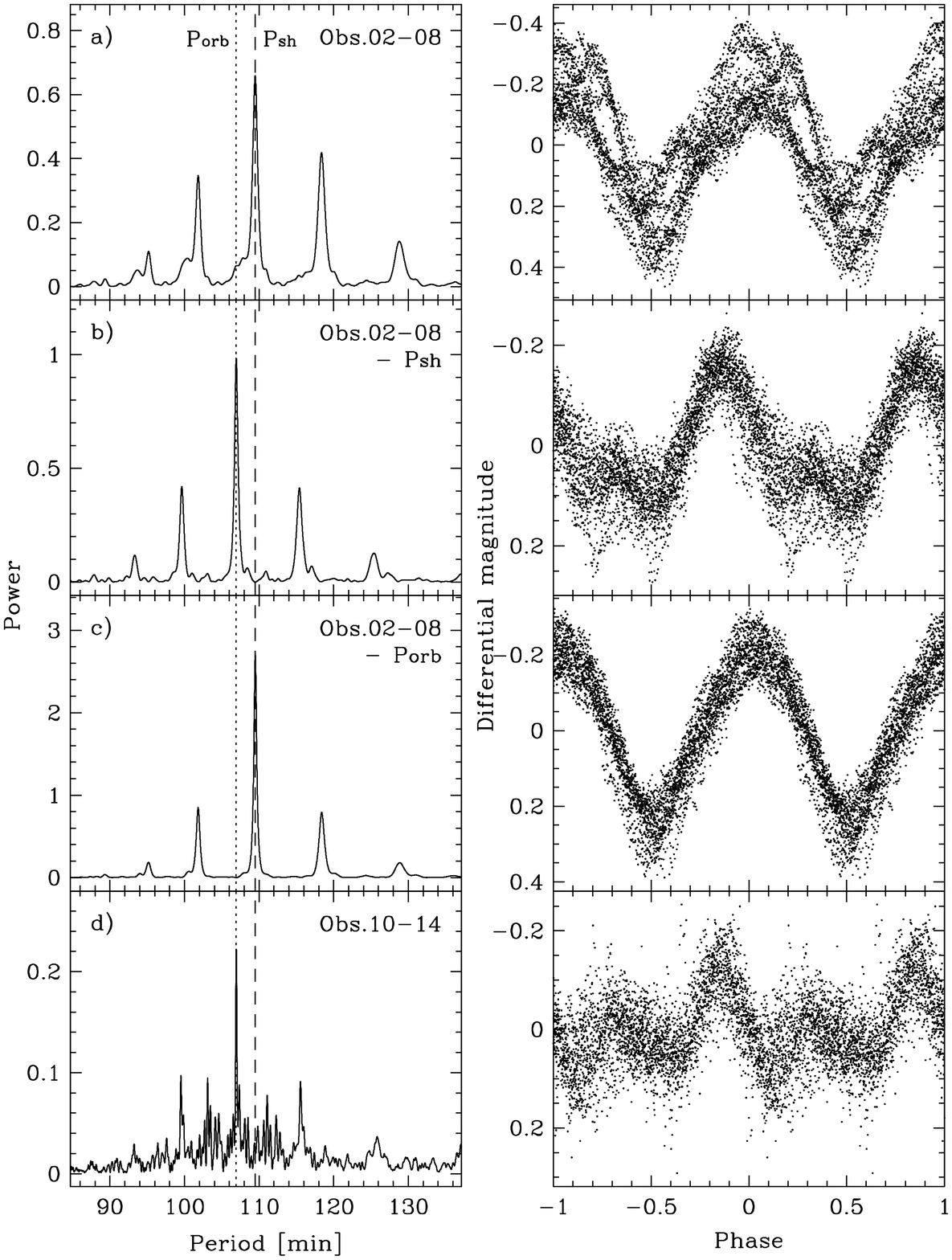}

\plotone{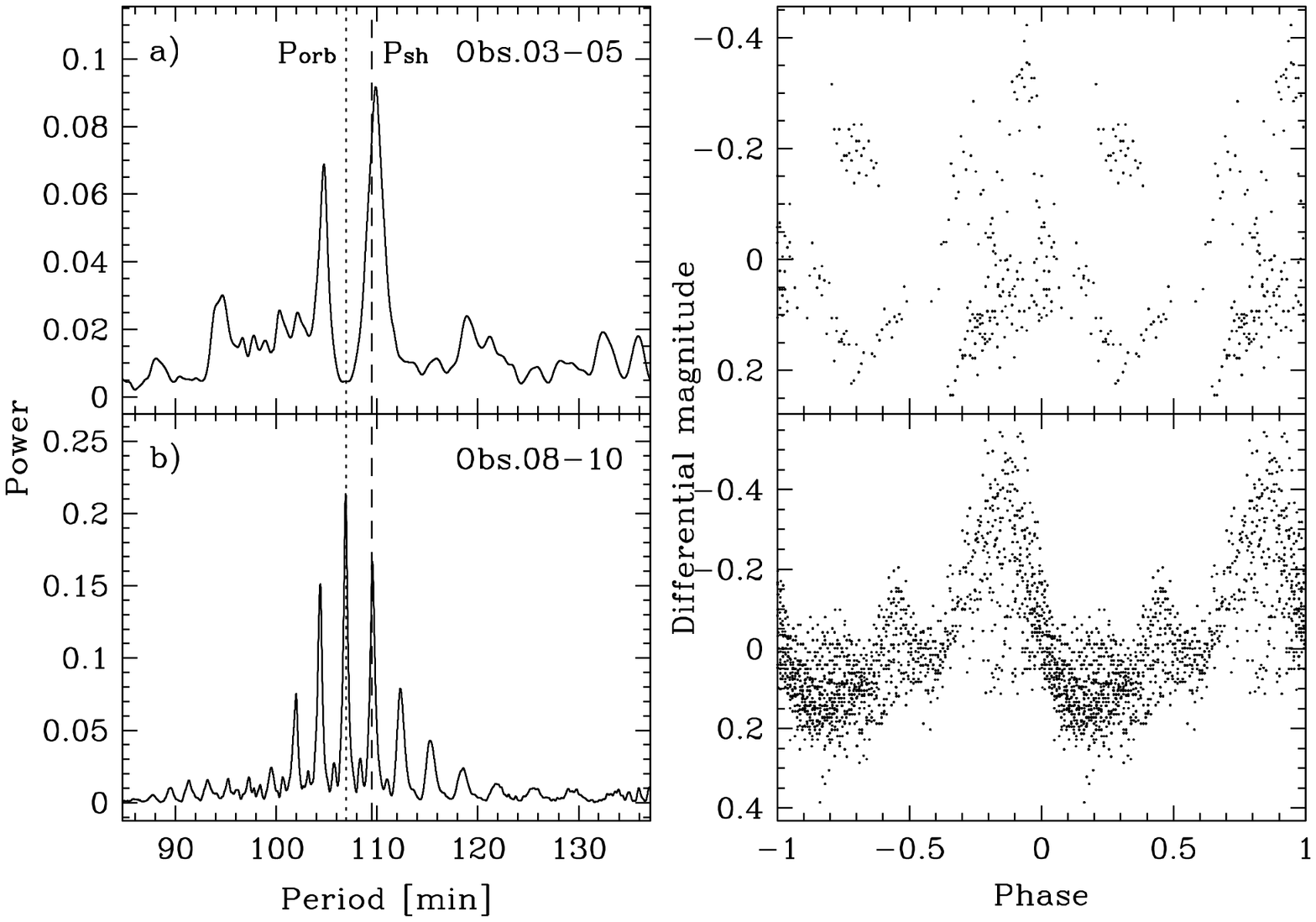}

\plotone{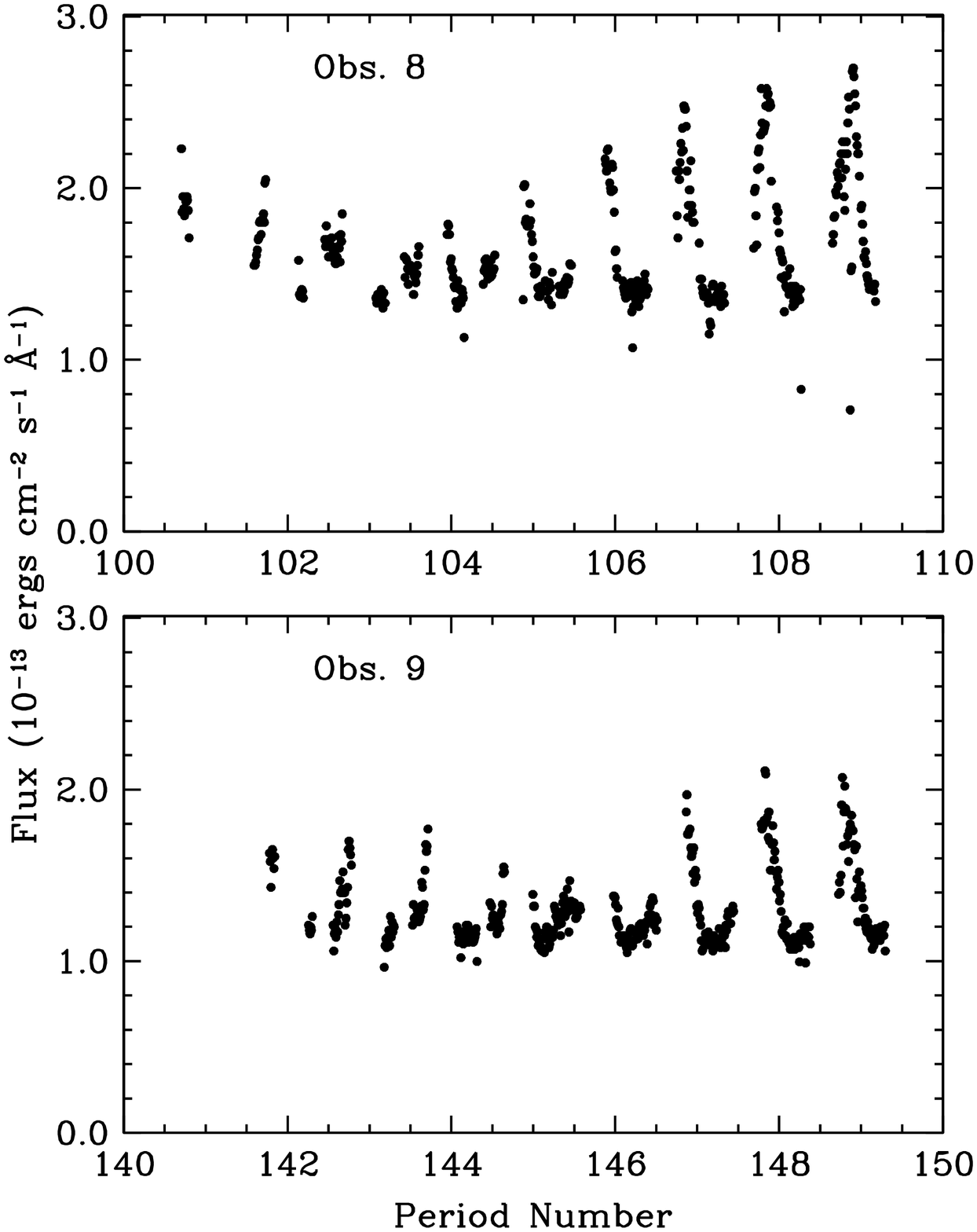}

\plotone{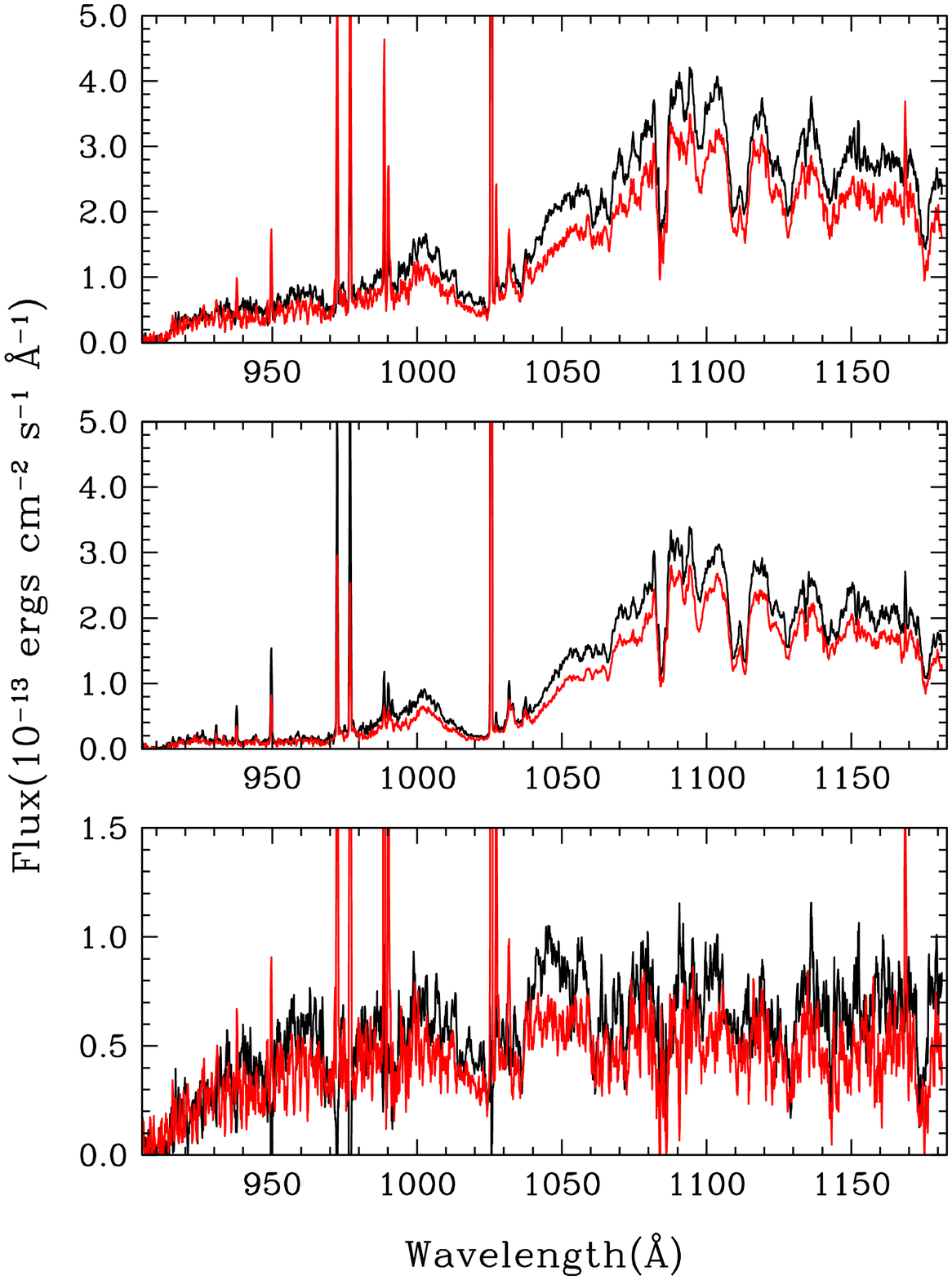}

\plotone{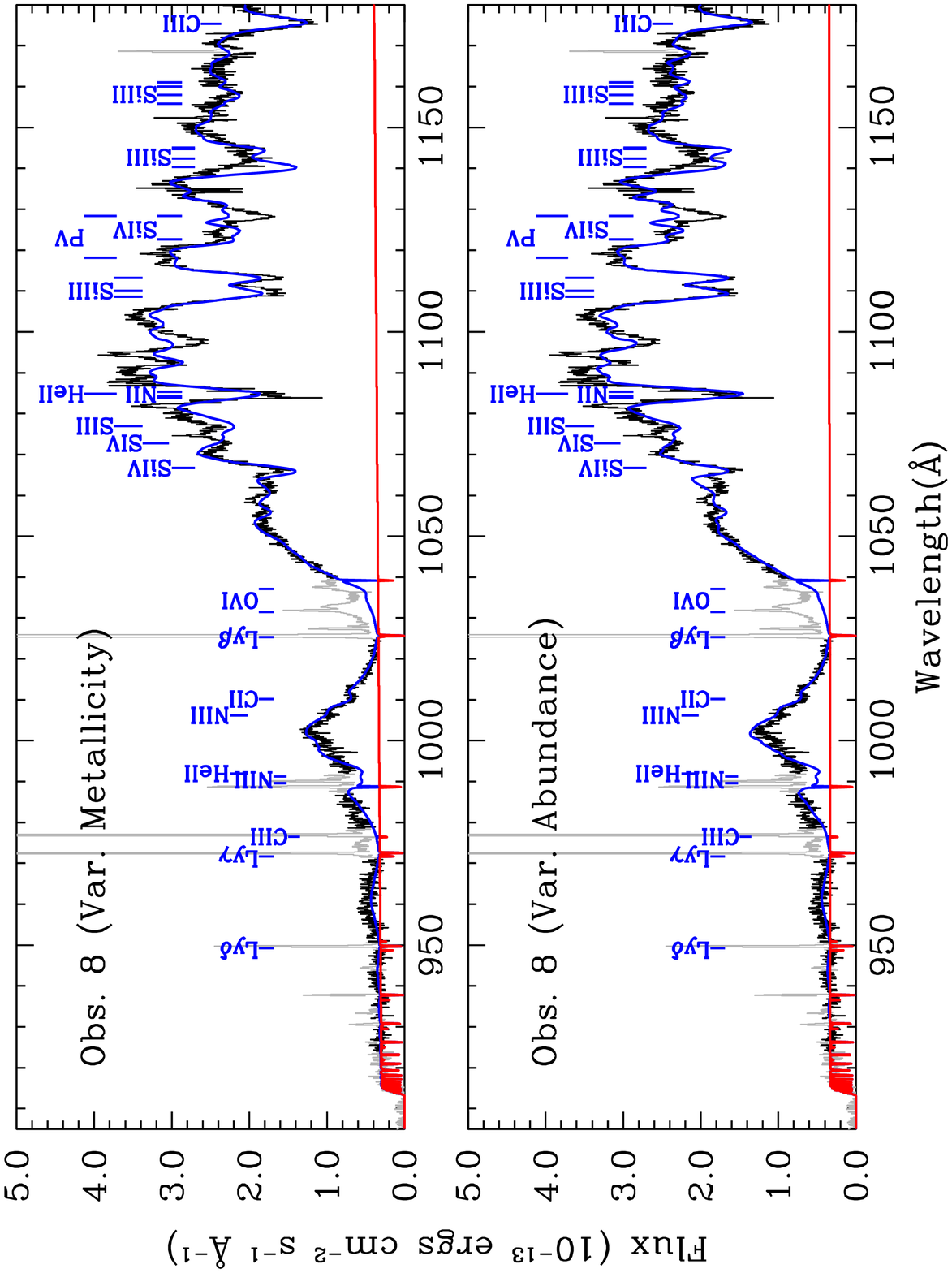}

\plotone{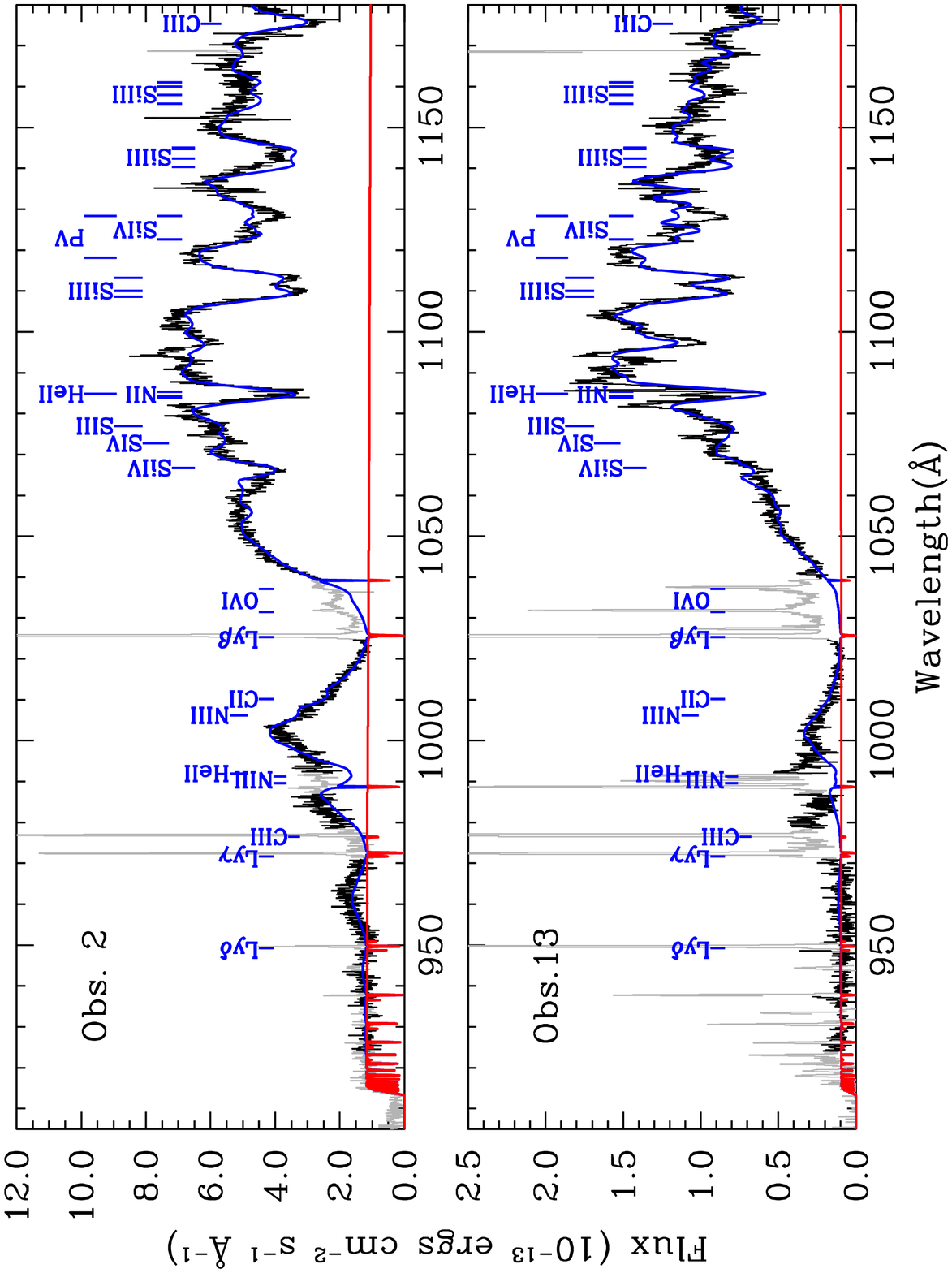}

\plotone{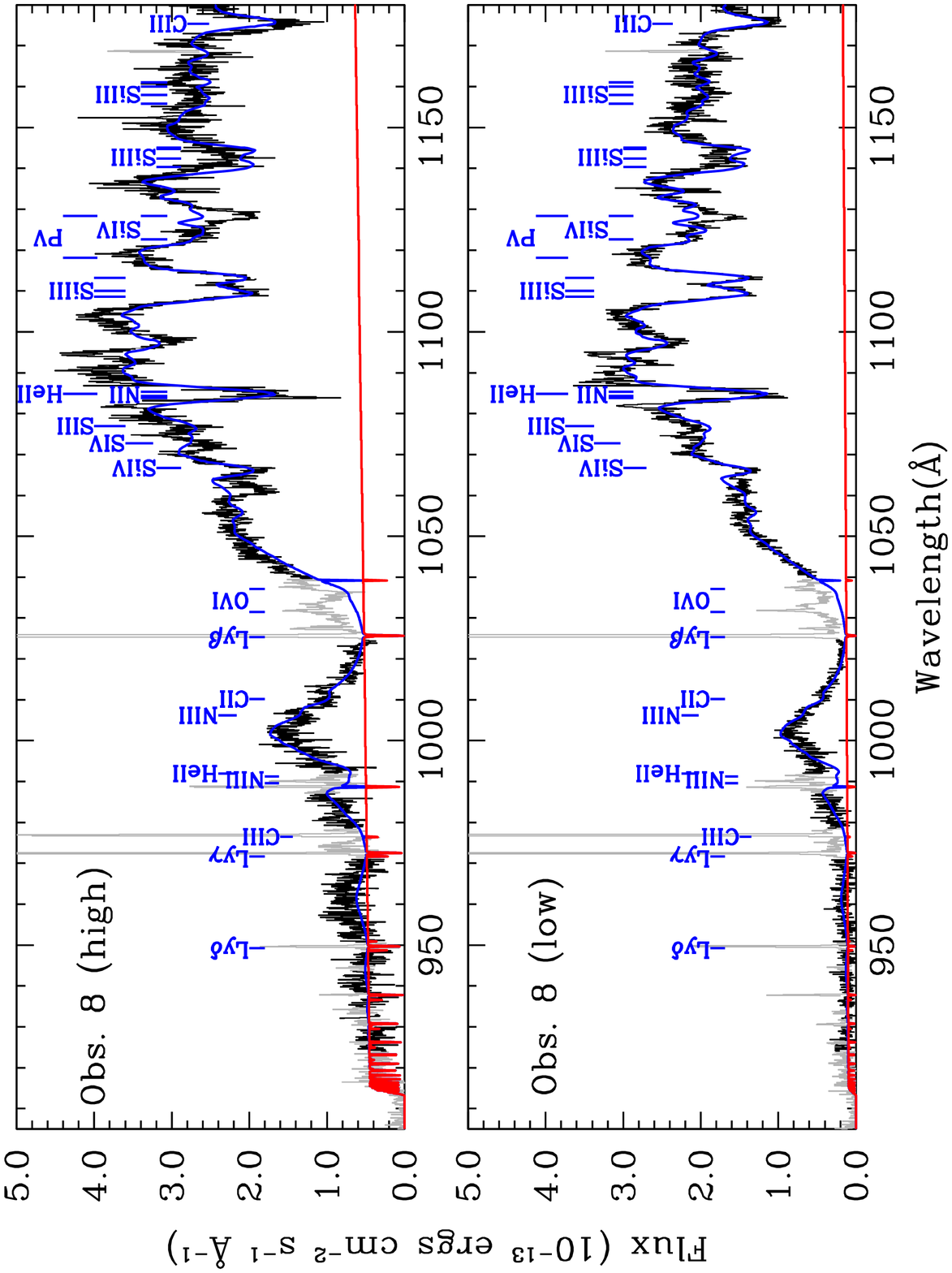}

\plotone{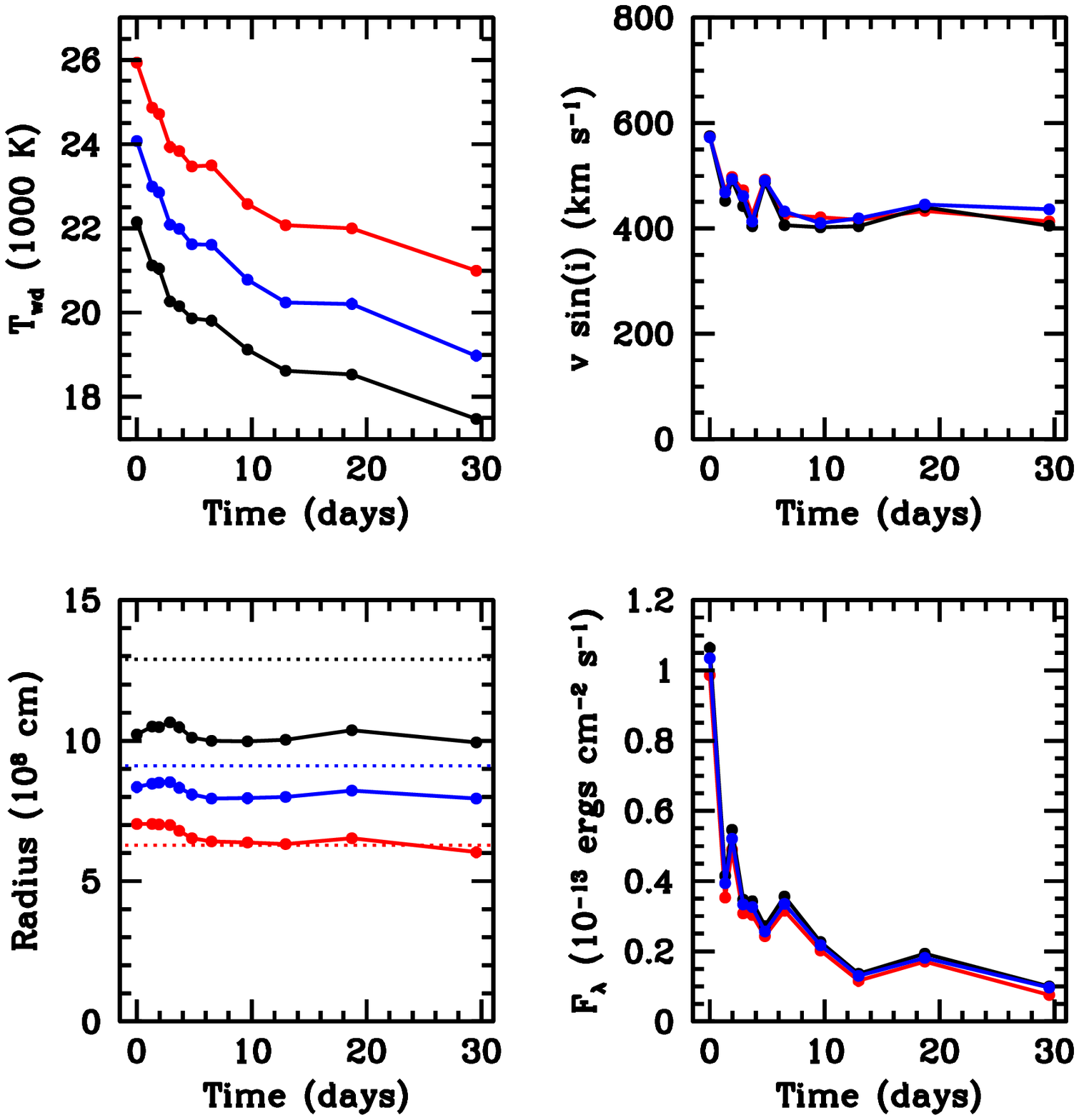}

\plotone{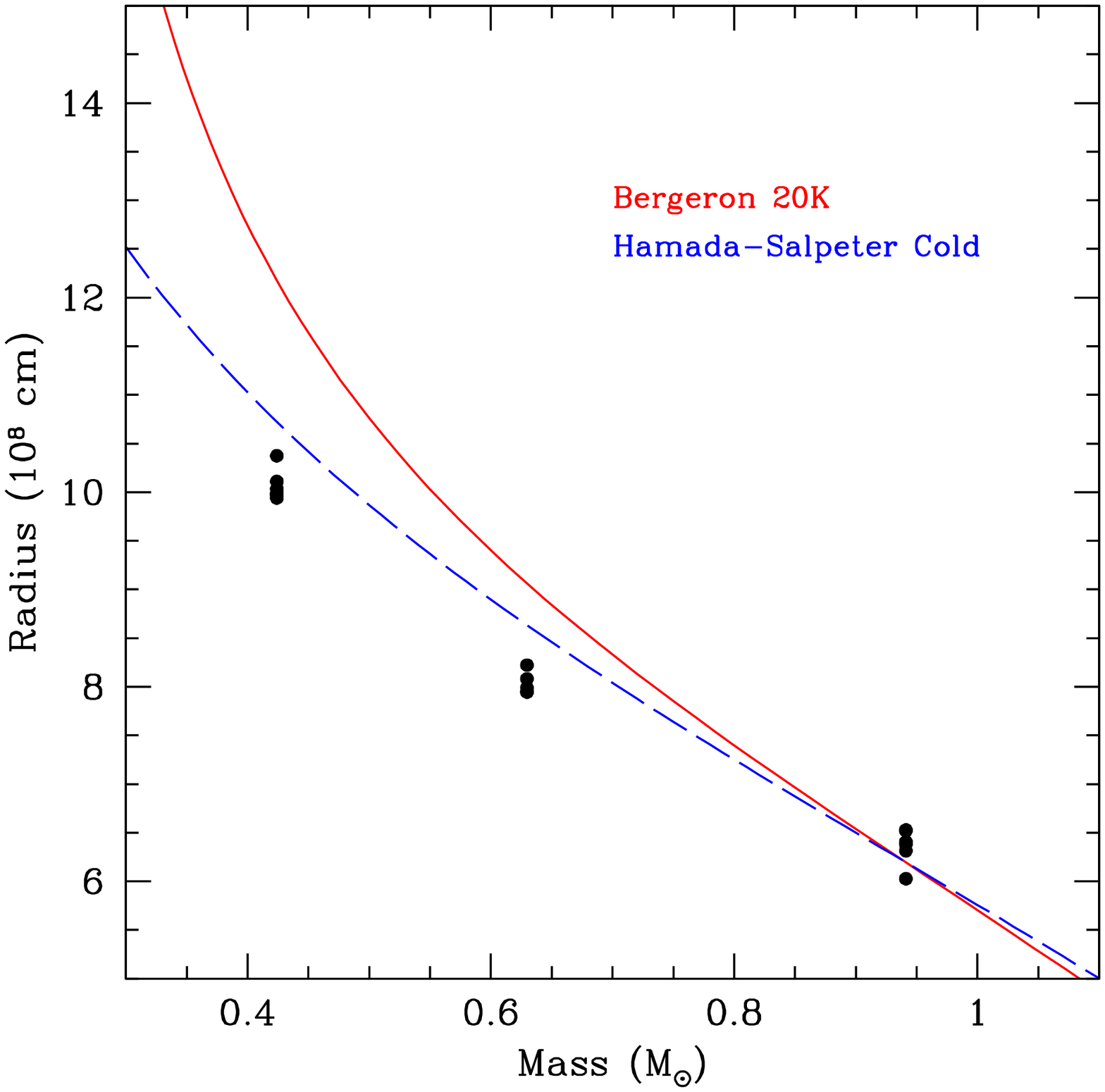}


\plotone{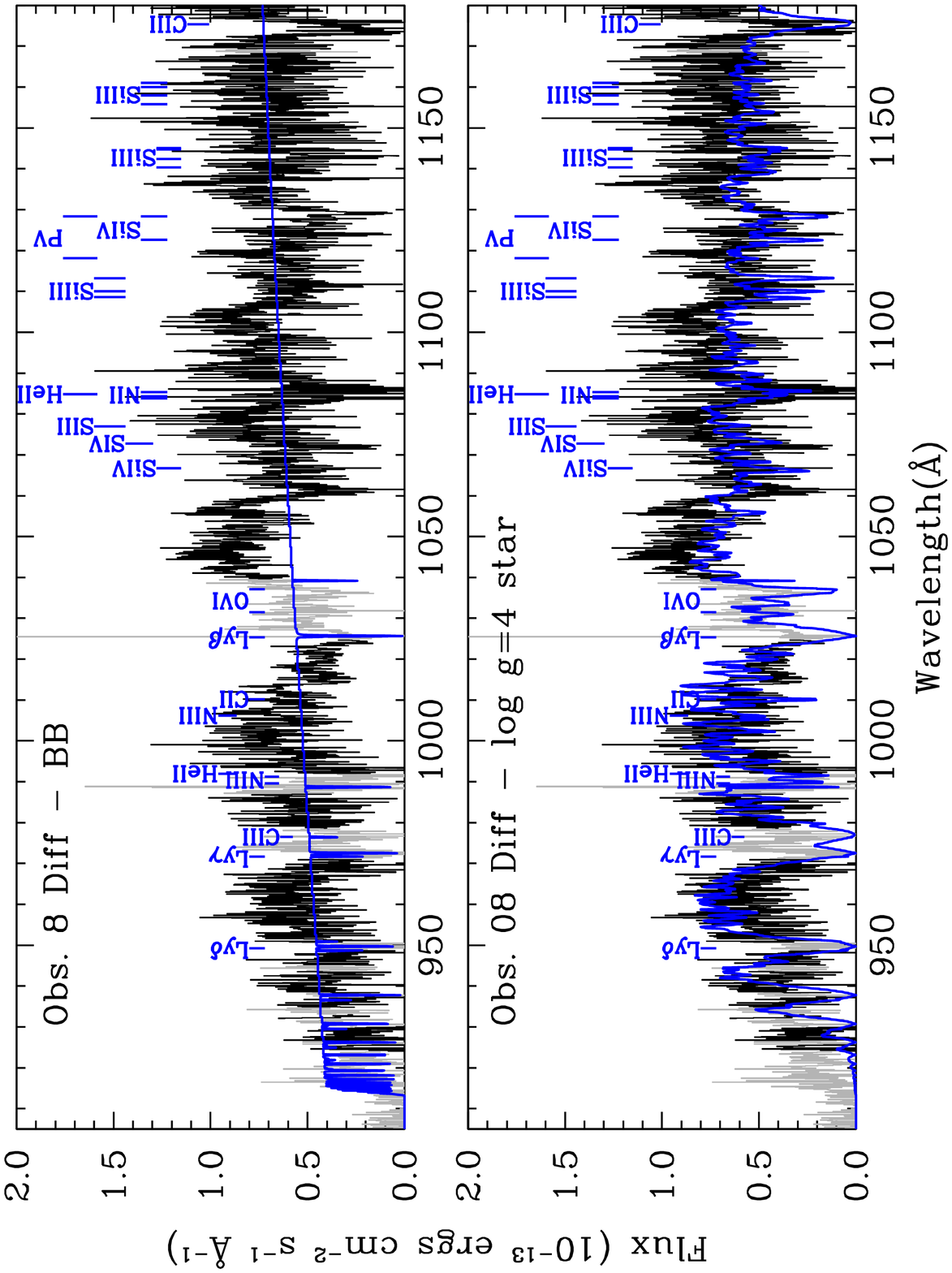}


\plotone{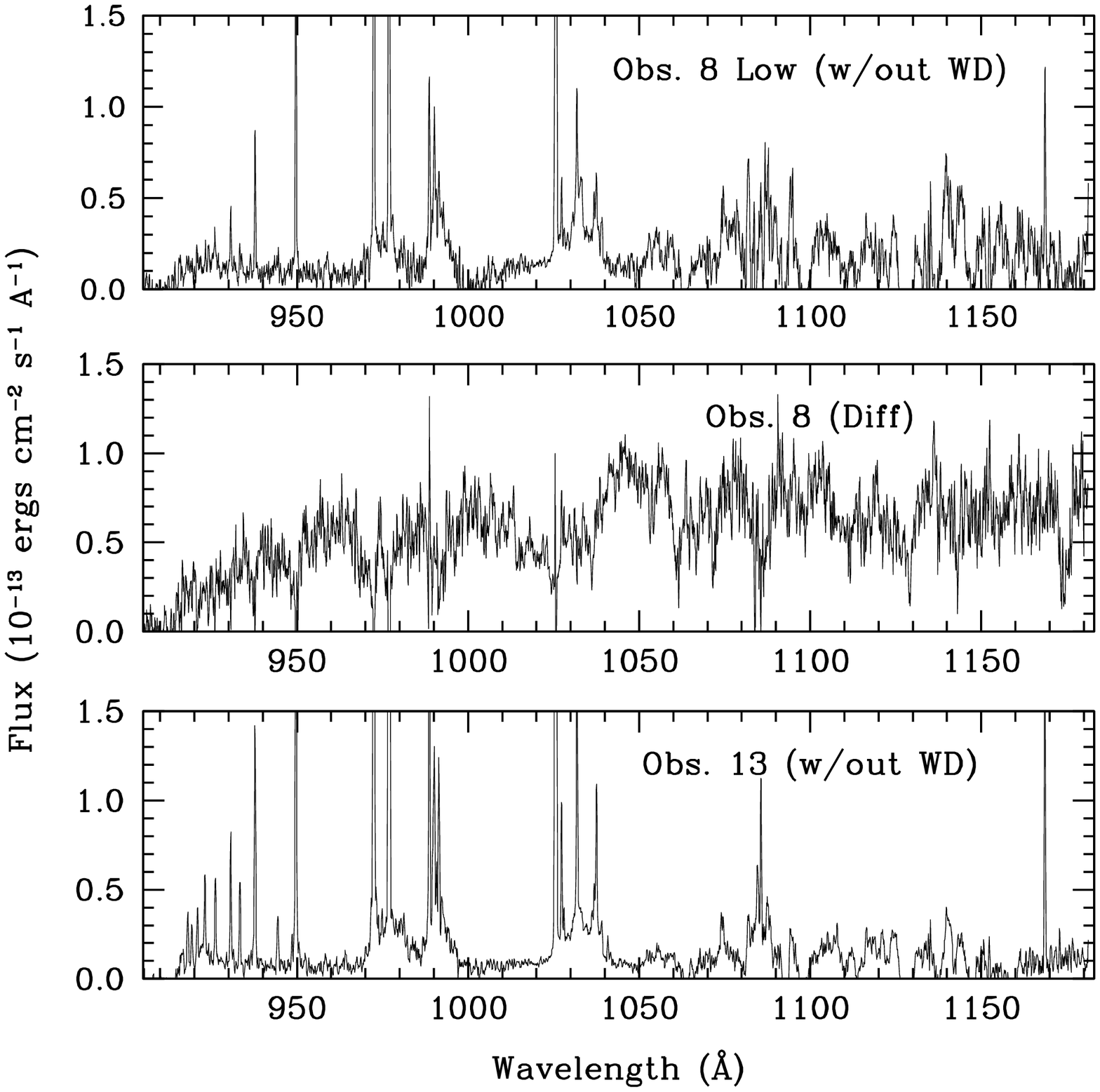}

\plotone{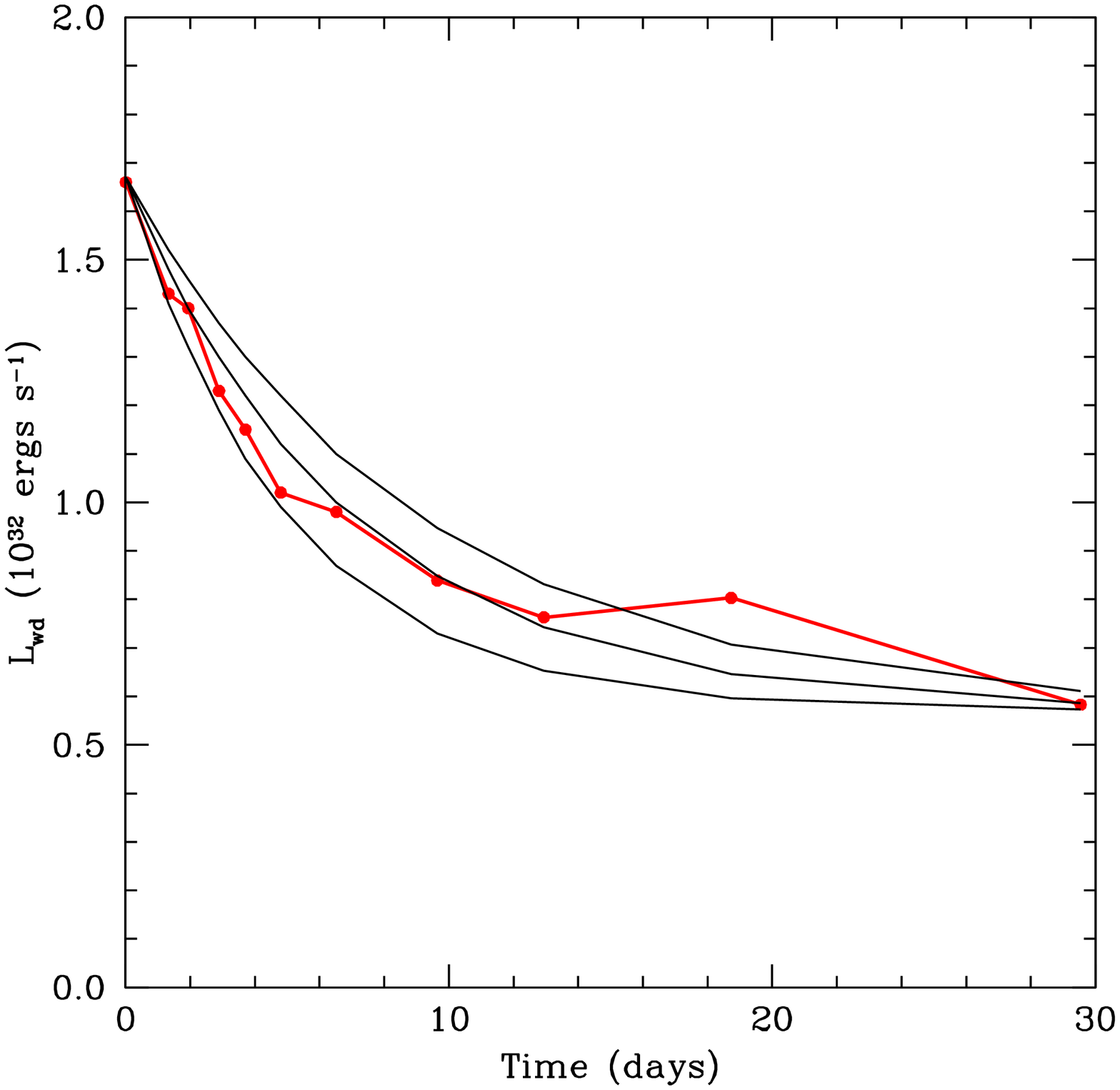}


\begin{thebibliography}{}

\bibitem[Armitage \& Livio(1996)]{Armitage1996} Armitage, P.~J., \& Livio, M.\ 1996, \apj, 470, 1024 



\bibitem[Araujo-Betancor et 
al.(2005)]{Araujo2005} Araujo-Betancor, S., et al.\ 2005, \aap, 430,
629  



\bibitem[Bergeron et al.(2001)]{Bergeron2001} Bergeron, P., Leggett, S.~K., \& Ruiz, M.~T.\ 2001, \apjs, 133, 413 


\bibitem[Dixon et al.(2007)]{Dixon2007} Dixon, W.~V., et al.\ 
2007, \pasp, 119, 527 


\bibitem[G{\"a}nsicke \& Beuermann(1996)]{Gaensicke1996} G{\"a}nsicke, B.~T., \& Beuermann, K.\ 1996, \aap, 309, L47 

\bibitem[G{\"a}nsicke et al.(2003)]{Gaensicke2003} G{\"a}nsicke, B.~T., et al.\ 2003, \apj, 594, 443 

\bibitem[G{\"a}nsicke et al.(2005)]{Gaensicke2005} G{\"a}nsicke, B.~T., Szkody, P., Howell, S.~B., 
\& Sion, E.~M.\ 2005, \apj, 629, 451 

\bibitem[Godard (2004)]{Godard2004}
Godard, B. 2004, ``IDF Cookbook'' 

\bibitem[Godon \& Sion(2005)]{Godon2005} Godon, P., \& Sion, E.~M.\ 2005, \mnras, 361, 809 

 Hubble Space Telescope STIS Spectroscopy and Modeling of the Long-Term Cooling of WZ Sagittae following the 2001 July Outburst  
\bibitem[Godon et al.(2006)]{Godon2006} Godon, P., Sion, E.~M., Cheng, F., Long, K.~S., G{\"a}nsicke, B.~T., 
\& Szkody, P.\ 2006, \apj, 642, 1018 



\bibitem[Godon et al.(2004)]{Godon2004} Godon, P., Sion, E.~M., Cheng, F.~H., Szkody, P., Long, K.~S., \& Froning, C.~S.\ 
2004, \apj, 612, 429 

\bibitem[Hamada \& Salpeter(1961)]{hamada1961} Hamada, T.~\&
Salpeter, E.~E.\ 1961, \apj, 134, 683

\bibitem[Harrison et al.(2005a)]{Harrison2005a} Harrison, T.~E., Howell, S.~B., Szkody, P., 
\& Cordova, F.~A.\ 2005a, \apjl, 632, L123 

\bibitem[Harrison et al.(2004)]{Harrison2004} Harrison, T.~E., Johnson, J.~J., McArthur, B.~E., Benedict, G.~F., Szkody, P., 
Howell, S.~B., \& Gelino, D.~M.\ 2004, \aj, 127, 460 


\bibitem[Harrison et al.(2005b)]{Harrison2005b} Harrison, T.~E., Osborne, H.~L., \& Howell, S.~B.\ 2005b, \aj, 129, 2400 


\bibitem[Hubeny(1988)]{Hubeny1988}
Hubeny, I. 1988, Comput. Phys. Comm., 52, 103


\bibitem[Hubeny \& Lanz(1995)]{Hubeny1995} Hubeny, I.~\& Lanz, T.\
1995, \apj, 439, 875



\bibitem[Knigge(2006)]{Knigge2006} Knigge, C.\ 2006, \mnras, 373, 484 


\bibitem[Ko et al.(1996)]{Ko1996} Ko, Y.-K., Lee, Y.~P., Schlegel, E.~M., \& Kallman, T.~R.\ 1996, \apj, 457, 363 



\bibitem[Kromer et al.(2007)]{Kromer2007} Kromer, M., Nagel, T., \& Werner, K.\ 2007, \aap, 475, 301 

\bibitem[Liu et al.(2008)]{Liu2008} Liu, F.~K., Meyer, F., Meyer-Hofmeister, E., \& Burwitz, V.\ 2008, \aap, 483, 231 

\bibitem[Long et al.(1993)]{Long1993} Long, K.~S., Blair, W.~P., Bowers, C.~W., Davidsen, A.~F., Kriss, G.~A., Sion, 
E.~M., \& Hubeny, I.\ 1993, \apj, 405, 327 

\bibitem[Long et al.(1996)]{Long1996} Long, K.~S., Blair, W.~P., Hubeny, I., \& Raymond, J.~C.\ 1996, \apj, 466, 964 

\bibitem[Long et al.(2006)]{Long2006} Long, K.~S., Brammer, G., \& Froning, C.~S.\ 2006, \apj, 648, 541 

\bibitem[Long et al.(2005)]{Long2005} Long, K.~S., Froning, C.~S., Knigge, C., Blair, W.~P., Kallman, T.~R., \& Ko, 
Y.-K.\ 2005, \apj, 630, 511 






\bibitem[Marks \& Sarna(1998)]{Marks1998} Marks, P.~B., \& Sarna, M.~J.\ 1998, \mnras, 301, 699 



\bibitem[Mauche et al.(1997)]{Mauche1997} Mauche, C.~W., Lee, Y.~P., \& Kallman, T.~R.\ 1997, \apj, 477, 832 






\bibitem[Mennickent et al.(2004)]{Mennickent2004} Mennickent, R.~E., Diaz, M.~P., \& Tappert, C.\ 2004, \mnras, 347, 1180 

\bibitem[Merritt et al.(2007)]{Merritt2007} Merritt, J., Night, C., \& Sion, E.~M.\ 2007, \pasp, 119, 251 

\bibitem[Meyer et al.(2000)]{Meyer2000} Meyer, F., Liu, B.~F., \& Meyer-Hofmeister, E.\ 2000, \aap, 361, 175 

\bibitem[Mineshige et al.(1998)]{Mineshige1998} Mineshige, S., Liu, B., Meyer, F., 
\& Meyer-Hofmeister, E.\ 1998, \pasj, 50, L5 

\bibitem[Mohanty \& Schlegel(1995)]{Mohanty1995} Mohanty, P., \& Schlegel, E.~M.\ 1995, \apj, 449, 330 

\bibitem[Moos et al.(2000)]{Moos2000} Moos, H.~W., et al.\ 2000, \apjl, 538, L1 


\bibitem[Osaki(1989)]{Osaki1989} Osaki, Y.\ 1989, \pasj, 41, 1005 

\bibitem[Paquette et al.(1986)]{Paquette1986} Paquette, C., Pelletier, C., Fontaine, G., \& Michaud, G.\ 1986, \apjs, 61, 197 


\bibitem[Pandel et al.(2003)]{Pandel2003} Pandel, D., C{\'o}rdova, F.~A., \& Howell, S.~B.\ 2003, \mnras, 346, 1231 

\bibitem[Patterson(1998)]{Patterson1998a} Patterson, J.\ 1998, \pasp, 110, 1132 

\bibitem[Patterson et al.(1998)]{Patterson1998} Patterson, J., 
Richman, H., Kemp, J., \& Mukai, K.\ 1998, \pasp, 110, 403 


\bibitem[Polidan et al.(1990)]{Polidan1990} Polidan, R.~S., Mauche, C.~W., \& Wade, R.~A.\ 1990, \apj, 356, 211 

\bibitem[Press et al.(2007)]{NumRecipes}
Press, W. H., Teukolsky, S. A., Vetterling,W. T., \& Flannery, B. P. 2007, Numerical Recipes:
the Art of Scientific Computing, 3rd edn. (Cambridge: Cambridge U. Press), 807-818

\bibitem[Piro et al.(2005)]{Piro2005} Piro, A.~L., Arras, P., \& Bildsten, L.\ 2005, \apj, 628, 401 

%
\bibitem[Rogoziecki \& Schwarzenberg-Czerny(2001)]{Rogoziecki2001}
Rogoziecki, P., \& Schwarzenberg-Czerny, A.\ 2001, \mnras, 323, 850  

\bibitem[Sahnow et al.(2000)]{Sahnow2000} Sahnow, D.~J., et al.\ 2000, \apjl, 538, L7 


\bibitem[Schoembs \& Vogt(1981)]{Schoembs1981} Schoembs, R., \& Vogt, N.\ 1981, \aap, 97, 185 

\bibitem[Schreiber et 
al.(2004)]{Schreiber2004} Schreiber, M.~R., Hameury, J.-M., \& Lasota,
J.-P.\ 2004, \aap, 427, 621

\bibitem[Schwarzenberg-Czerny(1996)]{Schwarzenberg1996} 
Schwarzenberg-Czerny, A.\ 1996, \apjl, 460, L107 


\bibitem[Shaviv \& Wehrse(1986)]{Shaviv1986} Shaviv, G., \& Wehrse, R.\ 1986, \aap, 159, L5 


\bibitem[Sion(1995)]{Sion1995} Sion, E.~M.\ 1995, \apj, 438, 876 


\bibitem[Sion et al.(2004)]{Sion2004} Sion, E.~M., Cheng, F.~H., G{\"a}nsicke, B.~T., \& Szkody, P.\ 2004, \apjl, 614, 
L61

\bibitem[Sion et al.(1997)]{Sion1997} Sion, E.~M., Cheng, F.~H., Sparks, W.~M., Szkody, P., Huang, M., \& Hubeny, I.\ 
1997, \apjl, 480, L17 


\bibitem[Sion et al.(2001)]{Sion2001} Sion, E.~M., Cheng, F.-H., Szkody, P., G{\"a}nsicke, B., Sparks, W.~M., 
\& Hubeny, I.\ 2001, \apjl, 561, L127 

\bibitem[Smith et al.(2006)]{Smith2006} Smith, A.~J., Haswell, C.~A., \& Hynes, R.~I.\ 2006, \mnras, 369, 1537 

\bibitem[Tappert et 
al.(2007)]{Tappert2007} Tappert, C., G{\"a}nsicke, B.~T., Schmidtobreick, L., Mennickent, R.~E., \& Navarrete, F.~P.\ 2007, \aap, 475, 575 
 
\bibitem[Townsley \& Bildsten(2003)]{Townsley2003} Townsley, D.~M., \& Bildsten, L.\ 2003, \apjl, 596, L227 


\bibitem[van Amerongen et al.(1987)]{VanAmerongen1987} van Amerongen, S., Damen, E., Groot, M., Kraakman, H., \& van Paradijs, 
J.\ 1987, \mnras, 225, 93 

\bibitem[van der Woerd \& Heise(1987)]{vanderWoerd1987} van der Woerd, H., \& Heise, J.\ 1987, \mnras, 225, 141 


\bibitem[van der Woerd et al.(1988)]{VanDerWoerd1987} van der Woerd, 
H., van der Klis, M., van Paradijs, J., Beuermann, K.,
\& Motch, C.\ 1988, \apj, 330, 911 



\bibitem[Verbunt et al.(1987)]{Verbunt1987} Verbunt, F., Hassall, B.~J.~M., Pringle, J.~E., Warner, B., \& Marang, F.\ 1987,
\mnras, 225, 113 

\bibitem[Vogt(1983)]{Vogt1983} Vogt, N.\ 1983, \aap, 118, 95 


\bibitem[Warner(1987)]{Warner1987} Warner, B.\ 1987, \mnras, 227, 23 

\bibitem[Warner(1995)]{Warner1995} Warner, B.\ 1995.\ Cataclysmic 
Variable Stars, Cambridge 
Astrophysics Series, Cambridge, New York: Cambridge University Press,  pp 81-82


\bibitem[Williams(1980)]{Williams1980} Williams, R.~E.\ 1980, \apj, 235, 939 

\end{thebibliography}
\end{document}